\pdfoutput=1
\documentclass[11pt]{article}

\usepackage[preprint]{acl}

\usepackage{times}
\usepackage{latexsym}
\usepackage{enumitem}
\usepackage{amsmath}
\usepackage{amssymb}
\usepackage{amsthm}

\usepackage[T1]{fontenc}
\usepackage[utf8]{inputenc}

\usepackage{microtype}

\usepackage{inconsolata}

\usepackage{tabularx}
\usepackage{graphicx}
\usepackage{xspace}
\usepackage{url}
\usepackage{hyperref}
\usepackage{booktabs}
\usepackage{listings}
\usepackage{xcolor}
\usepackage{tcolorbox}
\usepackage{lscape}
\usepackage{bm}

\newcommand\blfootnote[1]{%
  \begingroup
  \renewcommand\thefootnote{}\footnote{#1}%
  \addtocounter{footnote}{-1}%
  \endgroup
}

\definecolor{codegreen}{rgb}{0,0.6,0}
\definecolor{codeblue}{rgb}{0,0,0.6}
\definecolor{codegray}{rgb}{0.5,0.5,0.5}
\definecolor{backcolor}{rgb}{0.95,0.95,0.95}

\lstdefinestyle{pythonstyle}{
    backgroundcolor=\color{backcolor},   
    commentstyle=\color{codegreen},
    keywordstyle=\color{codeblue}\bfseries,
    numberstyle=\tiny\color{codegray},
    stringstyle=\color{codegreen},
    basicstyle=\ttfamily\footnotesize,
    breakatwhitespace=false,         
    breaklines=true,                 
    captionpos=b,                    
    keepspaces=true,                 
    numbers=left,                    
    numbersep=5pt,                  
    showspaces=false,                
    showstringspaces=false,
    showtabs=false,                  
    tabsize=4
}

\newcommand{\mQ}{\mathbf{Q}\xspace} % quality oracle
\newcommand{\mG}{\mathbf{G}\xspace} % perturbation graph
\newcommand{\mW}{\mathbf{W}\xspace} % watermarking scheme
\newcommand{\mP}{\mathbf{P}\xspace} % perturbation oracle
\newcommand{\mM}{\mathbf{M}\xspace} % watermarking model

\newcommand{\Vq}{\mathcal{V}_{x}^{\geq q}\xspace}
\newcommand{\Eq}{\mathcal{E}_{x}^{\geq q}\xspace}
\newcommand{\vertex}{\mathcal{V}_{x}\xspace}
\newcommand{\edge}{\mathcal{E}_{x}\xspace}
\newcommand{\epert}{\epsilon_{\text{pert}}\xspace}
\newcommand{\edist}{\epsilon_{\text{dist}}\xspace}
\theoremstyle{definition}
\newtheorem{definition}{Definition}[section]
\newtheorem{theorem}{Theorem}

\title{Sandcastles in the Storm: \\Revisiting the (Im)possibility of Strong Watermarking}

 \author{Fabrice Harel-Canada* \quad  Boran Erol* \quad Connor Choi \quad Jason Liu \quad Gary Jiarui Song \\ \quad  \quad \textbf{Nanyun Peng}  \quad \textbf{Amit Sahai}\\[7pt]
         University of California, Los Angeles\\[3pt]
         {
         \texttt{fabricehc@cs.ucla.edu}
         }
         }

\begin{document}
\maketitle

\begin{abstract}
Watermarking AI-generated text is critical for combating misuse. Yet recent theoretical work argues that any watermark can be erased via random walk attacks that perturb text while preserving quality. However, such attacks rely on two key assumptions: (1) rapid mixing (watermarks dissolve quickly under perturbations) and (2) reliable quality preservation (automated quality oracles perfectly guide edits). Through large-scale experiments and human-validated assessments, we find \textbf{mixing is slow}: 100\% of perturbed texts retain traces of their origin after hundreds of edits, defying rapid mixing. \textbf{Oracles falter}, as state-of-the-art quality detectors misjudge edits (77\% accuracy), compounding errors during attacks. Ultimately, \textbf{attacks underperform}: automated walks remove watermarks just 26\% of the time -- dropping to 10\% under human quality review. These findings challenge the inevitability of watermark removal. Instead, practical barriers -- slow mixing and imperfect quality control -- reveal watermarking to be far more robust than theoretical models suggest. The gap between idealized attacks and real-world feasibility underscores the need for stronger watermarking methods and more realistic attack models.
\end{abstract}

\blfootnote{* Equal contribution.}
\section{Introduction}

% \cite{publiclydetectable} Publicly detectable "weak" watermarking scheme that mentions the strong impossibility of watermarking.

% \cite{difficult_image} Very recent paper by the that shares an author with the one above (thought this might be relevant) on image watermarking that also mentions WITS.

% \cite{sok} SoK that mentions WITS as a paraphrasing attack.

% \cite{second_KGW_paper} On the Reliability of Watermarks for Large Language Models. Mentions that the sandpaper attack fails for "long enough" watermarked text, citing the sandpaper. Dismisses it pretty quickly.

% Link to Key Assumptions throughout the paper?

The rapid proliferation of generative AI has created an urgent need for mechanisms to authenticate machine-generated content. Watermarking -- embedding statistical signals into AI outputs to verify provenance -- serves as a vital safeguard against misinformation, IP theft, and academic fraud. While traditional methods employ visual patterns (e.g., pixel-level changes in images), statistical watermarking for text encodes imperceptible signals at lexical or semantic levels through specially selected patterns of tokens \cite{wmsurvey}. However, recent work by \citet{wits} (``Watermarks in the Sand,'' WITS) challenges the viability of watermarking, asserting that any such scheme can be defeated without degrading output quality through a simple random walk attack (see also, e.g.,~\citet{reliability, robust, dipper}). This impossibility result threatens to undermine the accountability and security of generative AI, leaving no viable path to enforce ethical standards or trace misuse.

The text-based WITS attack employs two primary components: (1) a perturbation oracle $ \mP $  that iteratively modifies text, and (2) a quality oracle $ \mQ $ to ensure that the edits are reasonable. These induce a random walk on a (potentially enormous) graph $ \mG $, where nodes represent possible texts $y$ and edges denote size-bounded perturbations (e.g., single-word swaps or paraphrases). Under certain assumptions, the random walk converges to a stationary distribution -- a stable equilibrium over nodes that remains unchanged under further perturbations. Crucially, this stationary distribution is a function of $ \mP $ and therefore independent of any particular watermarking scheme. As the random walk approaches this equilibrium, the likelihood of encountering a $ \mQ $-approved unwatermarked text increases. Notably, the WITS attack prioritizes quality equivalence over semantic equivalence: it seeks unwatermarked texts that score similarly under $ \mQ $ , even if their meaning diverges significantly from the original. 

While elegant in theory, the WITS argument relies on two key assumptions (KA) that warrant further scrutiny. Specifically, WITS assumes that:

\begin{itemize}[align=left, labelwidth=0.5cm, labelsep=0.15cm, leftmargin=0.75cm, noitemsep, nolistsep]  
    \item[\textbf{KA1.}] \textbf{The transition probabilities assigned to quality-preserving perturbations are high enough to ensure rapid mixing.} Formally, this means that the second-largest eigenvalue (in absolute value) of the transition matrix is sufficiently close to zero to ensure rapid mixing (\cite{wits}, Theorem 5).
    \item[\textbf{KA2.}] \textbf{The quality oracle $ \mQ  $ can reliably preserve output quality throughout the attack.} But if $ \mQ $ is unreliable -- either by admitting low-quality outputs or by blocking valid edits -- the attack either fails to escape the watermark or produces low-quality outputs that are no longer competitive with the original.
\end{itemize}

Taken together, \textbf{KA1} is concerned with attack efficiency and \textbf{KA2} further requires that the results remain meaningfully close to the initial text quality. To investigate whether these assumptions hold in practice, we designed analyses carefully tailored to study each assumption. For \textbf{KA1}, acquiring the eigenvalues of the transition matrix is infeasible due to its intractable size. Instead, we approximate mixing behavior by testing whether random walks retain memory of their starting states. If the random walk efficiently mixes, perturbed texts should lose memory of their starting points, making them indistinguishable from those originating elsewhere in the graph. Conversely, if stationary mixing is slow, initial states should remain identifiable even after many perturbations. 

For \textbf{KA2}, we crafted a dataset of perturbations annotated with human quality judgments and benchmarked a variety of automated oracles to determine their reliability. We then used the best oracle to guide the random-walk attacks and cross-checked the quality of the final perturbed texts to fairly estimate the robustness of several representative watermarking schemes -- KGW \cite{kgw}, SIR \cite{sir}, and Adaptive \cite{adaptive}. Our approach therefore addresses three primary research questions:

\begin{itemize}[align=left, labelwidth=0.5cm, labelsep=0.15cm, leftmargin=0.75cm, noitemsep, nolistsep] 
    \item[\textbf{RQ1.}] \emph{Can stationary distributions for watermarking be reached under practical constraints?} Even after hundreds of perturbations, starting states remain 100\% distinguishable, strongly suggesting that stationary distributions are not within efficient reach.
    \item[\textbf{RQ2.}] \emph{Are LLM-based quality oracles sophisticated enough to guide a random-walk attack?} The top-performing oracle attained an F1-score of 77.4\%, leaving significant room for errors to accumulate during the attack. This suggests that current generative oracles do not conform to the widely held belief that ``verification is easier than generation.''
    \item[\textbf{RQ3.}] \emph{How effective are random-walk attacks in breaking watermarks when controlling for quality?} Our improved random-walk attacks -- whether operating on a word, span, sentence, or document level -- succeeded in erasing the watermarks only 26.1\% of the time on average. After humans reviewed the perturbed texts to determine if quality was truly preserved, success dropped to an average 10.5\%.
\end{itemize}

Overall, our findings demonstrate a disconnect between theoretical assumptions and practical realities.  These findings highlight the trade-offs adversaries face: preserving quality necessitates minimal edits, but escaping detection requires riskier perturbations that compromise output quality. By bridging theoretical critique with empirical validation, this work challenges the inevitability of strong watermarking’s failure and offers a path forward for developing robust watermarking techniques grounded in real-world constraints.
\section{Background}
\label{label:background}

% % FHC Notes:
% % We should use this section to introduce:
% % - more formal definitions of P, Q, and G
% % - relevance of to perturbation step size and prompt entropy level

In this section, we outline the main objects and assumptions that underpin our analysis, following \cite{wits}. For formal definitions and more details, refer to Appendix~\ref{appendix:formal-definitions}.

Let $ \mM $ be a generative model mapping prompts $ x \in \mathcal{X}$ to outputs $ y \in \mathcal{Y}$ according to a probability distribution. Let $\mQ: \mathcal{X} \times \mathcal{Y} \to [0,1]$ be a function that returns a quality score for $ y $ as a response to prompt $ x $. We assume that the adversary has oracle access to $\mQ$. Notice that the watermarked model can be used as the quality oracle since we are not editing $ y $ using $\mQ$, whether or not this is sufficient to approximate $ \mQ $ is the content of \textbf{KA2}. 

Let $\mP: \mathcal{X}\times\mathcal{Y} \to \mathcal{Y}$ be a randomized \emph{perturbation oracle} that generates an alternative response $y'$ from an original response $y$ for the same prompt $x$. For the attack to succeed, $\mP$ must preserve the quality of $y$ with constant nonzero probability $ \epert \in (0,1]$ (Definition \ref{def:perturbation-preserving}).

Starting from a watermarked response \( y_0 \), we iteratively apply \( \mP \) to generate mutations by setting $ y_i = \mP(x, y_{i-1}) $.
To maintain high quality, each mutation must satisfy \( \mQ(x, y_i) \ge q \) (for some \( q \in [0,1] \)); otherwise, it is rejected.

We now formalize the graph $\mG_{x}^{\geq q}$ underlying the random walk induced by this process as the graph whose nodes are the output space of $\mM$ when given $x$ as input such that $\mQ(x,y) \geq q$, and whose edges are all pairs $(y,y')$ such that $\Pr\bigl[y' = \mP(x,y)\bigr] > 0$, with the weight of the edge given by $\Pr\bigl[y' = \mP(x,y)\bigr]$ (Definition~\ref{def:graph_representation}).

To ensure the success of the WITS attack, we need to impose mixing assumptions on the random walk, \emph{irreducibility} (Definition $ \ref{def:irreducibility}$) and \emph{aperiodicity} (Definition $ \ref{def:aperiodic}$). Together, these assumptions ensure that the random walk converges to a unique stationary distribution $ \vec{\pi} $ (Definition \ref{def:stationary-distribution}). In particular, after a sufficient number of steps, the probability of being at any node becomes independent of the initial state. This is critical for the WITS attack analysis: irreducibility guarantees that the random walk is not trapped within a single connected component, and aperiodicity prevents cyclic behavior that could hinder convergence. Given the enormous size of \( \mG \), aperiodicity is expected to hold. We discuss the irreducibility assumption further in Section~\ref{discussion}.

We now define the mixing time of an irreducible and aperiodic graph:

\begin{definition} [\cite{wits}, Definition 9]
    Let \( \mG = (\mathcal{V}, \mathcal{E}) \) be an irreducible and aperiodic weighted directed graph with transition matrix \( \vec{P} \). For any \( \edist \in (0,1] \), the \(\edist\)-\textit{mixing time} \( t_{\min} (\edist) \) of \( \vec{P} \) is the smallest \( t \) such that for every starting distribution \( \mathbf{p}_0 \in \mathbb{R}^{n} \), we have
    \[
    \left| \mathbf{p}_t - \vec{\pi} \right| = \left| (\vec{P}^{\top})^t \cdot \mathbf{p}_0 - \vec{\pi} \right| \leq \edist,
    \]
    where $ \mathbf{p}_t $ denotes the distribution over the vertices after $ t $ steps.
\end{definition}

After \( t_{\min}(\edist) \) steps, with probability at least \( 1 - \edist \), a sample drawn from the random walk behaves as if drawn from the stationary distribution -- i.e. independent of the original watermarked text.

Moreover, the mixing time \( t_{\min}(\edist) \) can be bounded in terms of the second largest eigenvalue \( g \) (in absolute value) of \( \vec{P} \) and the minimum stationary probability \( \pi_{\min} = \min\{\vec{\pi}(1), \dots, \vec{\pi}(n)\} \) 
\[
t_{\min}(\edist) \leq O\!\left( \frac{1}{1 - g} \cdot \log \!\left( \frac{1}{\pi_{\min} \cdot \edist} \right) \right).
\]
In practice, particularly for prompts with high entropy where the number of acceptable outputs (and hence the size of \( \vec{P} \)) is extremely large, estimating \( g \) and thus \( t_{\min}(\edist) \) becomes challenging. This difficulty directly relates to \textbf{KA1} and underscores the adversary’s challenge in determining when to halt the random walk. This is discussed further in Appendix \ref{appendix:attack_inefficiency}.

It is important to note that for an attack to be considered successful, the adversary $A$ must be significantly weaker than the model $\mM$. Otherwise, $A$ could simply ignore the watermarked output $ y $ and generate a fresh answer to $ x $, thereby trivially bypassing the watermark. Also notice that the step size of $ \mP $ directly impacts the mixing time of the random walk, which motivates the choice of our perturbation oracles.

At a high level, Theorem 2 in \cite{wits} proves that if these mixing conditions are satisfied, the random walk attack breaks any watermarking scheme with running time proportional to $ \frac{1}{1 - g} $. Moreover, the attacker can control the trade-off between quality of the final unwatermarked text and the probability of removing the watermark. % (Theorem \ref{theorem:wits_main_theorem} in Appendix).
\section{Evaluation Setup}
\label{section:evaluation_setup}

We now describe the main components of our evaluation: the watermarking schemes, the dataset, the automated quality metrics, and the perturbation oracles. We defer quality-oracle details to RQ2, where we benchmark and justify using \texttt{InternLM} as our primary $\mQ$ in our attacks.

\paragraph{Watermarkers.} We evaluate three widely used watermarking schemes $\mW$: KGW~\cite{kgw}, SIR~\cite{sir}, and Adaptive~\cite{adaptive}. Each embeds signals into generated text to enable authorship attribution. KGW utilizes a ``red-green'' list of tokens determined by the rolling hash of the previous $k$ tokens (typically $k=3$ to $5$). The logit scores for ``green'' tokens are boosted slightly to promote their selection. SIR follows a similar structure but instead relies on the semantic embeddings of preceding tokens, making it a form of ``semantic'' watermarking. Adaptive restricts its modifications to tokens in high-entropy regions to preserve text quality while still embedding a watermark. Because SIR and Adaptive each incorporate semantic context, both qualify as semantic watermarking schemes designed to resist attacks that preserve meaning through paraphrase. We note that these watermarking schemes produce detection scores on different scales: some, such as Adaptive, use a 0--100 scale, whereas KGW and SIR compute a $z$-statistic. Additional details about the watermarkers can be found in Appendix \ref{appendix:watermark_details}.

\paragraph{Dataset.} As noted in Section \ref{label:background}, the number of valid responses to a prompt (i.e., its \emph{entropy}) influences the structure of the perturbation graph $\mG$. To systematically investigate the relationship between entropy and attack success, we constructed a dataset for \textbf{RQ1} and \textbf{RQ3} featuring entropy-controlled prompts in three domains relevant to authorial accountability: education, journalism, and creative writing. 

For each domain, we designed a series of prompts with increasing specificity. For instance, a broad request might be ``Write a 500-word news article,'' while a more constrained one could read ``Write a 500-word news article about a global climate summit''(see Appendix \ref{appendix:entropy_prompts} for more details). We used the \texttt{Llama-3.1-70B-Instruct} model (denoted as $\mM$) to generate three watermarked texts per prompt for each of the three watermarking schemes, resulting in 270 watermarked texts. To provide a baseline, we also generated 90 unwatermarked texts from the same model. In addition, we included unwatermarked outputs from \texttt{GPT-4o} to represent scenarios with higher quality text that adversaries might favor. For each watermarking scheme, we computed the mean watermark detection score and its standard deviation on the unwatermarked texts to establish a reference range against which we measure whether a perturbed text remains distinguishable.

\paragraph{Automatic Quality Metrics.} We automatically evaluated text quality with several metrics. First, we used \texttt{InternLM}~\cite{internlm} as a reward model to acquire a numerical quality score. We used GPT-2~\cite{gpt2} to calculate perplexity, with lower values indicating higher fluency and predictability. In addition, we counted the number of grammatical errors using a standard grammar checker~\cite{language_tool}, and we computed the unique bigrams in each text to assess lexical diversity. Table~\ref{table:watermark_dataset}  in Appendix~\ref{appendix:dataset_statistics} summarizes these metrics for both unwatermarked and watermarked texts before attack.

\paragraph{Perturbation Oracles} Our perturbation oracles, denoted as $\mP$, apply adversarial edits at different levels of granularity. \texttt{WordMutator} and \texttt{EntropyWordMutator} replace individual tokens, with the latter concentrating on high-entropy tokens that are most likely to carry stronger watermark signals. \texttt{SpanMutator}, which is the primary text-based perturbation approach used by \citet{wits}, masks $n=6$ contiguous tokens and refills them using \texttt{T5-XL}~\cite{t5}. \texttt{SentenceMutator} paraphrases a single sentence at each step with \texttt{Llama-3.1-8B}, while \texttt{DocumentMutator}, \texttt{Document1StepMutator}, and \texttt{Document2StepMutator} attempt a full-document paraphrase, either in one pass or in multiple passes. We allow a sufficient number of steps (e.g., 1000 for token-level edits and 100 for document-level edits) to maximize the opportunity for watermark removal. Further technical details on each $\mP$ are provided in Appendix~\ref{appendix:perturbation_oracles}.

\section{Analysis}

We now empirically examine the WITS assumptions by asking: (RQ1) whether the random-walk attacks reach a stationary distribution, (RQ2) whether LLM-based quality oracles reliably guide the attack, and (RQ3) how often watermark removal preserves text quality.

\subsection{RQ1. Can Stationary Distributions Be Reached Under Practical Constraints?}

\renewcommand{\arraystretch}{0.9}
\begin{table*}[!htbp]
\centering
\resizebox{0.8\textwidth}{!}{%
\begin{tabular}{lrrrrr}
\toprule
\textbf{$\mP$ Oracle}           & \textbf{Steps}                & \textbf{Tests}                & \textbf{\texttt{Llama-3.1-70B}} & \textbf{\texttt{GPT-4o}} & \textbf{\texttt{o3-mini-high}}   \\
\midrule
\texttt{Word}                & 1000                          & 720                           & 0                               & 0                        & 0                \\
\texttt{EntropyWord}         & 1000                          & 720                           & 0                               & 0                        & 0                \\
\texttt{Span}                & 250                           & 720                           & 12                              & 1                        & 0                \\
\texttt{Sentence}            & 150                           & 720                           & 38                              & 3                        & 0                \\
\texttt{Document}            & 100                           & 421                           & 2                               & 0                        & 0                \\
\texttt{Document1Step}       & 100                           & 576                           & 0                               & 0                        & 0                \\
\texttt{Document2Step}       & 100                           & 678                           & 1                               & 0                        & 0                \\
\midrule
\textbf{Total / Failed Tests}       & \multicolumn{1}{l}{\textbf{}} & \textbf{4555}                 & \textbf{53}                     & \textbf{4}               & \textbf{0}       \\
\midrule
\textbf{Cumulative Distinguished (\%)} & \multicolumn{1}{l}{\textbf{}} & \multicolumn{1}{l}{\textbf{}} & \textbf{98.84}                  & \textbf{99.91}           & \textbf{100}     \\
\bottomrule
\end{tabular}%
}
\vspace{-2mm}
\caption{Summary of failed distinguisher tests per $\mP$, along with the step budget and total tests. Classification is first performed by \texttt{Llama-3.1-70B}, followed by \texttt{GPT-4o} on its failures, then \texttt{o3-mini-high} on any remaining cases. The overall 100\% success rate indicates that the attacked texts never lose memory of their starting points, contradicting \textbf{KA1} and suggesting that a stationary distribution is not reached in practice.}
\vspace{-4mm}
\label{table:distinguisher}
\end{table*}

WITS posits that repeated perturbations rapidly decouple a text from its starting state, eventually sampling from a stationary distribution. If mixing is slow, however, watermark removal becomes impractical in real-world applications. Although the second-largest eigenvalue ($ g $) of the transition matrix provides a formal measure of mixing speed, computing $ g $ directly is infeasible due to the high dimensionality of $\mG$. Instead, we measure whether the final texts can be traced back to their initial forms. If this tracing remains accurate even after many perturbations, it suggests that the random walk has not mixed sufficiently.

To approximate mixing, we propose a novel \emph{lineage distinguisher} test. First, we choose two initial responses for each prompt. We then run a random-walk attack, perturbing one starting text until a $\mP$-specific step budget is used up (e.g., 1000 steps for \texttt{WordMutator}). Periodically sampled texts along this walk are then classified by \texttt{Llama-3.1-70B-Instruct}, which attempts to identify their true origin. Since well-mixed texts should be indistinguishable from random samples in $\mG$, classification accuracy should collapse to chance if a stationary distribution is reached.

\subsubsection{Results}

Table~\ref{table:distinguisher} summarizes results of a multi-stage classification approach designed to balance accuracy and cost. We first use \texttt{Llama-3.1-70B-Instruct} with a zero-shot prompt in a best-of-2 (see Appendix~\ref{appendix:distinguisher_prompt} for details). If \texttt{Llama-3} produces a tied result (considered a failure), we escalate to the stronger (but more expensive) \texttt{GPT-4o} \cite{gpt4o}. Any remaining cases are then passed to \texttt{o3-mini-high} \cite{o3mini}. At no point were both trials wrong in a best-of-2. Across all tests, \texttt{Llama-3.1-70B-Instruct} alone achieves 98.84\% accuracy. \texttt{GPT-4o} correctly resolves nearly all of the remaining 53 failures, and \texttt{o3-mini-high} succeeds on the last four, yielding a final 100\% success rate. This consistently high distinguishability shows that random walks do not adequately mix within the allotted steps, thus contradicting \textbf{KA1} and indicating that the attacked texts remain too similar to their originals for watermark removal to rely on a converged stationary distribution.

\subsection{RQ2. Are LLM-based quality oracles sophisticated enough to guide a random-walk attack?} 

A core assumption of WITS-style attacks is that verifying output quality is at least as easy as generating content (\cite{wits}, \S 4.1.2). This assumption aligns with the common belief that recognizing high-quality work -- whether in music, cinema, or literature -- is simpler than creating it. However, this premise has not been rigorously tested in the context of generative LLMs. If $\mQ$ is unreliable -- either by approving degraded outputs or blocking valid transformations -- the attack stalls or yields low-quality text. To examine this assumption systematically, we built and benchmarked a variety of LLM-based oracles, measuring their ability to preserve quality while guiding watermark removal.

\paragraph{The Sandcastles Benchmark.} We created the Sandcastles dataset to evaluate oracle reliability by sampling 100 diverse prompts from \texttt{arena-human-preference-55k}~\cite{chatbotarena}, generating watermarked responses, and applying up to 20 iterative perturbations. At the 1st, 10th, and 20th steps, we collected human annotations comparing the perturbed text to its original. To ensure unbiased evaluation, annotators were presented with two texts, A and B, without knowing which had been perturbed. They provided ternary preference judgments, selecting either A, B, or tie. 

\renewcommand{\arraystretch}{0.9}
\begin{table*}[!htbp]
\centering
\resizebox{0.8\textwidth}{!}{%
\begin{tabular}{lllrr}
\toprule
\textbf{Oracle}     & \textbf{Model}                                                                                                        & \textbf{QP Prec.} & \textbf{Overall F1}\\
\midrule
MutationOracle      & \href{https://huggingface.co/meta-llama/Llama-3.1-70B-Instruct}{Llama-3.1-70B-Instruct} \cite{llama3}                 & \textbf{84.62}        & 66.93          \\
Prometheus2Absolute & GPT-4-Turbo \cite{gpt4turbo}                                                                                          & 76.15                 & 67.55          \\
InternLMOracle      & \href{https://huggingface.co/internlm/internlm2-20b-reward}{internlm2-20b-reward} \cite{internlm}                     & 65.69                 & 69.84          \\
DiffOracle          & \href{https://huggingface.co/meta-llama/Llama-3.1-70B-Instruct}{Llama-3.1-70B-Instruct} \cite{llama3}                 & 71.74                 & 70.85          \\
DiffOracle+FT       & \href{https://huggingface.co/meta-llama/Llama-3.1-70B-Instruct}{Llama-3.1-70B-Instruct} + Fine-tuning                 & 69.07                 & 76.94          \\
MutationOracle+FT   & GPT-4o \cite{gpt4o} + Fine-tuning                                                                                     & 74.51                 & \textbf{77.38} \\
\bottomrule
\end{tabular}%
}
\vspace{-2mm}
\caption{Performance of selected quality oracles on human-annotated data (full results in Appendix~\ref{appendix:oracles}). QP Precision measures accuracy in preserving high-quality outputs, while Overall F1 reflects general classification performance. Despite fine-tuning, no oracle fully aligns with human judgments, challenging \textbf{KA2} and limiting their reliability in guiding random-walk attacks.}
\vspace{-4mm}
\label{table:oracles}
\end{table*}

For oracle training and evaluation, we binarized judgments: preferences for the perturbed text or a tie were labeled as "Quality Preserved," while preferences for the original were labeled as "Degraded." This simplification provides a clearer evaluation signal while preserving human preference patterns. The final dataset includes 795 annotated perturbations, with additional statistics in Appendix~\ref{appendix:sandcastle_details}.

\paragraph{Constructing and Evaluating Oracles.} As a baseline, we followed the WITS suggestion to reuse the watermarking model $\mM$ (\texttt{Llama-3.1-70B-Instruct}) as a quality oracle. After initial trials revealed positional biases and inconsistencies with human judgments, we explored several improvements. We ran oracle queries multiple times with flipped text orders, explicitly explained that the task involved assessing mutation quality (\texttt{MutationOracle}), and supplemented prompts with a changelog of all edits (\texttt{DiffOracle}). We then fine-tuned the strongest of these oracles on the Sandcastles training set (\texttt{MutationOracle+FT}, \texttt{DiffOracle+FT}). In parallel, we evaluated six reward models from the RewardBench leaderboard,\footnote{\url{https://huggingface.co/spaces/allenai/reward-bench}} each producing continuous scores that we thresholded (e.g., a 0.46 deviation from the original in \texttt{InternLMOracle}) to classify outputs as high-quality or degraded. Finally, even though cost concerns make large proprietary models impractical for full attacks, we tested \texttt{GPT-4-Turbo}, \texttt{GPT-4o}, and a fine-tuned version \texttt{GPT-4o+FT} to gauge whether more powerful models offer significant improvements. Additional details on these oracle variants appear in Appendix~\ref{appendix:oracles}.

We report both Quality Preserved (QP) Precision and Overall F1 to assess oracle performance. High QP Precision reduces false-positive approvals of degraded texts, a critical safeguard against cumulative quality erosion during multiple perturbations. The Overall F1 captures an oracle’s overall ability to classify text quality accurately.

\subsubsection{Results}

Table~\ref{table:oracles} summarizes each oracle’s runtime, return type, and performance. Our results show that current LLM-based quality oracles remain inconsistent, limiting the feasibility of using them to guide watermark removal attacks. Even the best-performing oracle (\texttt{GPT-4o+FT}) attains an Overall F1 of only 77.4\%, implying that nearly one in five perturbations is misclassified. Fine-tuning and the use of powerful models like \texttt{GPT-4o} and \texttt{GPT-4-Turbo} do reduce errors somewhat, but not to a level sufficient for reliably guiding multi-step attacks. This misclassification compounds over repeated perturbations, forcing adversaries to either accept noticeable quality loss or proceed with low attack efficiency.

Among locally hosted models, the most robust approaches used difference-aware or mutation-aware prompts -- \texttt{DiffOracle} (QP Precision: 70.9\%) and \texttt{MutationOracle} (QP F1: 66.9\%). Even after fine-tuning, however, these oracles still frequently labeled degraded outputs as high-quality. Moreover, high-scoring reward models from RewardBench (e.g., \texttt{INFORMOracle} at 95.1, \texttt{SkyworkOracle} at 94.3) often performed worse than simpler approaches, suggesting that generic reward tuning does not align well with the nuances of watermark-focused attacks. Collectively, these errors highlight a key limitation: LLM-based verification is not as reliable as assumed. We discuss potential causes for this limitations in Section~\ref{discussion}.

\paragraph{Extended Comparison.}

While \texttt{GPT-4o+FT} achieves the best results overall, its high cost makes it impractical for many-step attacks. We therefore ran a human evaluation comparing two locally hosted oracles -- the best boolean-based (\texttt{DiffOracle+FT}) and the best floating-point (\texttt{InternLMOracle}) -- in a 150-step attack using \texttt{SentenceMutator}, which induced the most mixing in \textbf{RQ1}. Human judges found that \texttt{InternLMOracle} preserved quality in 47.78\% of samples, compared to 40.0\% for \texttt{DiffOracle+FT} (Table~\ref{table:head2head}). Bayesian analysis indicated an 85.08\% probability that \texttt{InternLMOracle} is genuinely superior (Appendix~\ref{appendix:i_vs_d}), leading us to select \texttt{InternLMOracle} for further experiments despite its remaining error rate. 

\subsection{RQ3. How effective are random walk attacks in breaking watermarks when controlling for quality?}

% Use https://www.tablesgenerator.com/ to convert googlesheets table into LaTeX. 
\renewcommand{\arraystretch}{0.9}
\begin{table*}[!htbp]
\centering
\resizebox{0.85\textwidth}{!}{%
\begin{tabular}{llrrrrrrrrrrrrrrrr}
\toprule
\textbf{$\mW$} &
\textbf{$\mP$ Oracle} &
$\,\bm{\mu_{w_0}}$ &
$\,\bm{\mu_{w_t}}$ &
\textbf{$\text{ASR}_{\mathrm{min}}$} &
\textbf{$\text{ASR}_{\mathrm{fin}}$} &
\textbf{Reviewed} &
\textbf{QP} &
$\neg$\textbf{QP} &
\textbf{$\text{Q-ASR}_{\mathrm{fin}}$} \\
\midrule
\texttt{Adaptive} & \texttt{Word}          & 99.27 & 70.37 & 0.00  & 0.00  & 0  & 0  & 0  & 0.00  \\
\texttt{Adaptive} & \texttt{EntropyWord}   & 99.27 & 82.45 & 0.00  & 0.00  & 0  & 0  & 0  & 0.00  \\
\texttt{Adaptive} & \texttt{Span}          & 99.27 & 67.21 & 1.54  & 1.54  & 2  & 2  & 0  & 1.54  \\
\texttt{Adaptive} & \texttt{Sentence}      & 99.27 & 59.93 & 35.34 & 19.21 & 20 & 8  & 12 & 7.68  \\
\texttt{Adaptive} & \texttt{Document}      & 99.27 & 58.55 & 48.78 & 45.24 & 20 & 8  & 12 & 18.10 \\
\texttt{Adaptive} & \texttt{Document1Step} & 99.27 & 70.94 & 1.16  & 1.16  & 2  & 2  & 0  & 1.16  \\
\texttt{Adaptive} & \texttt{Document2Step} & 99.27 & 73.39 & 5.33  & 4.71  & 8  & 5  & 3  & 2.94  \\
\midrule
\texttt{SIR}      & \texttt{Word}          & 5.32  & 1.74  & 78.22 & 57.89 & 20 & 1  & 19 & 2.89  \\
\texttt{SIR}      & \texttt{EntropyWord}   & 5.32  & 3.30  & 39.68 & 27.54 & 20 & 0  & 20 & 0.00  \\
\texttt{SIR}      & \texttt{Span}          & 5.32  & 1.57  & 60.71 & 37.40 & 20 & 5  & 15 & 9.35  \\
\texttt{SIR}      & \texttt{Sentence}      & 5.32  & 0.52  & 87.65 & 74.71 & 20 & 13 & 7  & 48.56 \\
\texttt{SIR}      & \texttt{Document}      & 5.32  & 0.93  & 61.54 & 46.09 & 20 & 6  & 14 & 13.83 \\
\texttt{SIR}      & \texttt{Document1Step} & 5.32  & 2.54  & 14.04 & 14.04 & 12 & 11 & 1  & 12.87 \\
\texttt{SIR}      & \texttt{Document2Step} & 5.32  & 3.07  & 68.09 & 49.06 & 20 & 12 & 8  & 29.44 \\
\midrule
\texttt{KGW}      & \texttt{Word}          & 0.28  & 0.17  & 47.54 & 20.00 & 20 & 4  & 16 & 4.00  \\
\texttt{KGW}      & \texttt{EntropyWord}   & 0.28  & 0.22  & 3.45  & 0.56  & 1  & 0  & 1  & 0.00  \\
\texttt{KGW}      & \texttt{Span}          & 0.28  & 0.20  & 38.46 & 32.35 & 20 & 14 & 6  & 22.65 \\
\texttt{KGW}      & \texttt{Sentence}      & 0.28  & 0.14  & 89.47 & 56.52 & 20 & 7  & 13 & 19.78 \\
\texttt{KGW}      & \texttt{Document}      & 0.28  & 0.18  & 62.50 & 44.44 & 20 & 8  & 12 & 17.78 \\
\texttt{KGW}      & \texttt{Document1Step} & 0.28  & 0.27  & 12.66 & 8.54  & 14 & 7  & 7  & 4.27  \\
\texttt{KGW}      & \texttt{Document2Step} & 0.28  & 0.18  & 9.09  & 7.78  & 10 & 4  & 6  & 3.11  \\
\midrule
\multicolumn{4}{l}{\textbf{Averages (\%)}} & 36.44 & 26.13 &    & 40.48 & 59.52   & 10.47 \\
\bottomrule
\end{tabular}%
}
\vspace{-2mm}
\caption{Attack success rates (ASR) across different perturbation strategies. Human review reveals an average of 59.52\% of successfully attacked texts have degraded quality. $\bm{\mu_{w_0}}$ represents the initial watermark score at step 0, while $\bm{\mu_{w_t}}$ represents the final watermark score after $t$ mutation steps.``min'' refers to the point where the watermark score is at its lowest during the attack while ``fin'' refers to score at the final step of the attack. ``Reviewed'' indicates the number of human-reviewed examples where the watermark was broken. \textbf{QP} and $\neg$\textbf{QP} represent the number of cases where human reviewers judged the attacked text as quality-preserving or degraded, respectively. \textbf{Q-ASR}$_{\mathrm{fin}}$ is the re-estimated attack success after controling for quality, calculated as \textbf{ASR}$_{\mathrm{fin}} \times  (\text{QP} / \text{Reviewed})$.}
\vspace{-4mm}
\label{table:attack_results}
\end{table*}

% \caption{Attack success rates (ASR) across different perturbation strategies. Human review reveals an average of 59.52\% of successfully attacked texts have degraded quality. $\bm{\mu_{w_0}}$ represents the initial watermark score at step 0, while $\bm{\mu_{w_t}}$ represents the final watermark score after $t$ mutation steps.``min'' refers to the point where the watermark score is at its lowest during the attack while ``fin'' refers to score at the final step of the attack. "Reviewed" indicates the number of human-reviewed examples where the watermark was broken. \textbf{QP} and $\neg$\textbf{QP} represent the number of cases where human reviewers judged the attacked text as quality-preserving or degraded, respectively. \textbf{Q-ASR}$_{\mathrm{fin}}$ is the re-estimated attack success after controling for quality, calculated as \textbf{ASR}$_{\mathrm{fin}} \times  (\text{QP} / \text{Reviewed})$.}

\paragraph{Attack Methodology.} We apply various perturbation oracles to texts watermarked by KGW, SIR, or Adaptive. At each step, a candidate edit is proposed and accepted only if our quality oracle (\texttt{InternLMOracle}) labels it as high-quality. We track watermark detection scores and terminate when a fixed number of mutation steps is reached (details in Appendix~\ref{appendix:full_attack_details}). An attack is deemed successful if the final detection score is less than $\mu_{\text{uw}} + 2\sigma_{\text{uw}}$, where $\mu_{\text{uw}}$ and $\sigma_{\text{uw}}$ are, respectively, the mean and standard deviation of unwatermarked texts’ detection scores. Under the assumption that scores follow a normal distribution, being below this threshold places the text in a region where fewer than 2.3\% of unwatermarked samples lie above it, making it highly unlikely to be flagged as watermarked. We define the \emph{attack success rate} (ASR) as the proportion of final texts that satisfy this criterion. We record two key states along each attack trace: (1)
$s_{min}$, corresponding to the lowest watermark score achieved (as if an attacker had real-time detector feedback), and (2) $s_{fin}$, produced when the perturbation budget is exhausted without direct detector feedback. This distinction clarifies how close attacks can come to fully erasing the watermark under ideal versus practical conditions.

\noindent
\paragraph{Quality Evaluation.} 
Since automatic metrics can overlook subtle forms of degradation, we also ran a human evaluation using 10 volunteer annotators, primarily with STEM backgrounds, but no specialized linguistic expertise. Each annotator compared a watermarked text with its final attacked version, randomly labeled as \texttt{text\_A} or \texttt{text\_B}, and chose whether one was superior or if they were equivalent. Instructions (Appendix~\ref{appendix:human_study_details}) emphasized fluency, style, coherence, and prompt relevance. We gathered up to 20 successfully attacked texts per watermark-oracle pairing, then determined how often annotators judged the attacked text to be at least as good as the original. This proportion was used to adjust the baseline $\text{Q-ASR}_{\text{fin}}$ value, giving a more realistic estimate of attack success. 

\subsubsection{Results}

Table~\ref{table:attack_results} shows that watermarks remain highly resistant to random-walk attacks, especially under quality constraints. Across all perturbation strategies, final attack success rates (\textbf{ASR\textsubscript{fin}}) are low at 26.1\% on average, with Adaptive proving the most robust. After human review, adjusted success rates (\textbf{Q-ASR\textsubscript{fin}}) drop by an average of 59.52\%, with no fairly evaluated attack succeeding more than half the time. 

The effectiveness of attacks varies by perturbation type. Document-level edits perform best against Adaptive and KGW, while sentence-level perturbations show the highest success against SIR and KGW. Yet, even sentence-level mutations fail in most cases when quality is considered. Adaptive remains resistant to all strategies, especially entropy-controlled and word-level mutations, where no attack succeeds (\textbf{Q-ASR\textsubscript{min}} = 0.00). Word-level perturbations generally fail across all watermarking schemes.

Overall, these findings highlight a major limitation for adversaries: while watermark removal is theoretically possible, preserving quality remains a significant obstacle. Even under ideal conditions -- full perturbation budgets and access to detection feedback -- successful attacks often fail to maintain human-perceived text quality.
\section{Discussion}
\label{discussion}

\paragraph{The Tricky Irreducibility Assumption.}
Another core assumption underlying the WITS framework that is difficult to evaluate empirically is that $\mP$ induces an \emph{irreducible} graph $\mG^{\geq q}_x$. In other words, in theory, any high-quality text state is reachable from any starting point via a sequence of edits that all remain above the quality threshold $q$. However, this assumption is highly nontrivial, especially considering (a) the inherent limitations of $\mP$, (b) the fact that edits are often local, and (c) the possibility that some transitions may necessarily involve brief dips below the threshold.

To see why irreducibility might fail in practice, consider two high-quality responses to a prompt asking for a story: assume that one is Star Wars and another is The Lord of the Rings (LOTR). For one to transform into the other \emph{while remaining above the threshold}, there would need to be a sequence of high-quality intermediate texts that blend elements of both franchises. If our threshold $q$ is stringent -- say, requiring not just correct language but also stylistic consistency and thematic clarity -- then many “blend” stages would likely be muddled or incoherent, causing the text to fall below $q$.

Hence, it is reasonable to suspect that the high-quality subgraph might contain distinct “islands” that cannot reach one another without temporarily leaving $\mG^{\geq q}_x$. In fact, when humans write -- one character at a time -- they invariably pass through numerous low-quality states (partial words, half-formed sentences) before arriving at any one of the various ways of saying something of quality. Local edit operators, such as those that insert or delete single tokens or small chunks of text, face a similar risk: even a small disruption can degrade quality if the threshold is strict.

That said, irreducibility might still be recovered if we loosen our assumptions. For instance, we might allow momentary dips in quality during transitions so long as the process does not “get stuck” below $q$; or we could permit larger, more context-aware edits that can leap more cleanly between stylistic domains. In practice, these motivations led to the development of the \textbf{\texttt{Document2StepMutator}}, which aims to ensure that modifications are localized enough to avoid substantial quality degradation, yet also sufficiently broad to permit meaningful jumps. This design tries to strike a balance between remaining “near” high-quality states most of the time and retaining enough flexibility to move across different regions of the text space -- ideally preventing the formation of disconnected “islands” of high-quality text.

% Mention the connectivity metrics from the experiments? FHC: never got around to computing them and I don't think they'd be meaningful anyway. 

\paragraph{Why do LLMs Struggle to Verify?} While humans intuitively find verification easier than generation, this asymmetry may actually \emph{reverse} for LLMs due to their probabilistic architecture and training paradigms. The core tension arises from LLMs' design as next-token predictors \cite{brown2020language}, which optimizes them for fluency over factual accuracy or logical rigor \cite{bender2021dangers,lin2022truthfulqa}. Though techniques like chain-of-thought prompting \cite{wei2022chain} can simulate self-checking, the models remain fundamentally tuned to generate plausible continuations -- not to verify them. 

Compounding this, LLMs lack exposure to the iterative critique processes that shape human judgment. Trained on polished outputs \cite{dodge2021documenting}, they rarely encounter explicit revisions (e.g., drafts with margin notes like "this plot point contradicts Chapter 3") that teach cause-effect relationships between quality and text structure \cite{stiennon2020learning}. Consequently, their "critiques" often reduce to surface-level heuristics (e.g., associating complex syntax with professionalism) rather than principled reasoning.

Whether verification is inherently harder for LLMs may hinge on whether “quality” is reducible to “likelihood.” If not, their adeptness at generating fluent text may paradoxically hamper verification, as polished outputs mask subtle shortcomings \cite{bender2021dangers}, creating a hall-of-mirrors effect where plausibility is mistaken for truth.

% incorporate into next draft: https://www.sciencedirect.com/science/article/pii/S1364661324000275

% The difficulty of verification for LLMs ultimately hinges on whether “quality” can be reduced to “likelihood.” Current evidence suggests it cannot: models’ fluency paradoxically masks subtle errors \cite{bender2021dangers}, creating a hall-of-mirrors effect where generated text’s internal consistency is mistaken for factual reliability \cite{vankempen2024challenge}. Without architectural innovations or training paradigms that explicitly reward verification (e.g., reinforcement learning for error detection), this asymmetry will likely persist.
\section{Conclusion}

Our findings reveal that watermark removal via random-walk attacks is far less certain than theoretical work suggests. Slow mixing and imperfect quality verification create significant real-world barriers. These insights invite deeper investigation: evolving watermark schemes could exploit the difficulty of consistent, high-quality edits, while attackers must grapple with the costs and risks of large-scale text manipulation. Our study also highlights the need for quality measures that align with human judgment, not surface features. Addressing these challenges -- mixing speed, oracle reliability, and quality standards -- will ensure watermarking remains viable against sophisticated attacks.

\section*{Limitations}

While our findings highlight practical barriers to random-walk attacks, several limitations constrain their generalizability. First, we focus on three watermarking schemes (KGW, SIR, Adaptive) and specific perturbation oracles. Other schemes \cite{markllm, semamark} and attack methods, especially those with advanced error-correction or alternative pathways (e.g., \citet{rastogi-pruthi-2024-revisiting}), may yield different results. 

Second, while human verification is critical to assessing attack success -- a factor often overlooked in prior work -- our findings rely on a small, potentially non-representative group of annotators. Broader user studies, richer datasets, and more diverse oracle designs are needed to validate our conclusions across varied scenarios, though such efforts would require significant resources.

Third, our analyses rely on LLM-based oracles fine-tuned for quality judgment, which still misclassify ~20\% of edits. Future breakthroughs in text evaluation -- such as low-cost \emph{reasoning} models \cite{deepseek-r1} or specialized reward functions -- could improve verification accuracy to the levels required to sustain viable attacks. 

Finally, while we tested hundreds of perturbations, resource constraints limited exploration of arbitrarily large edit sequences. In theory, infinite steps might approach WITS’s stationary distribution, but our results reveal substantial practical barriers. Computational costs further hinder scalability: \texttt{DocumentMutator} (based on \texttt{DIPPER} \cite{dipper}) took 213 seconds per attack step, rendering large-scale edits impractical. %These runtimes also explain incomplete traces for some document-level attacks. 

% of course we cannot experiment with every mutator (and edit step-size)

% some of our initial texts have low quality

% final outputs of our mutators have formatting issues which could easily be fixed by a human, but we ignored them in the human annotations

% asterisk the adaptive doc mutator attacks, even though it's a full set. KGW and GPT doc mutator attacks also suck.

% adaptive doc mutators have 2 attacks with missing z-scores because it exceeded the ctx window.

\bibliography{custom}

\appendix

\onecolumn
\section{Appendix: Formal Definitions}
\label{appendix:formal-definitions}

In this section, we provide formal definitions of objects mentioned in Section $\ref{label:background}$ and elaborate on some definitions. As with the background section, most of these are directly from \cite{wits}. Let us begin by providing formal definitions of objects mentioned in Section \ref{label:background}.

\begin{definition}[\(\epert\)-Preserving Perturbation Oracle, \cite{wits}, Definition 6]
\label{def:perturbation-preserving}
Let \( \mP : \mathcal{X}\times\mathcal{Y} \to \mathcal{Y} \) be a randomized oracle that, given \( (x,y) \), outputs a new response \( y' \). The oracle \( \mP \) is said to be \(\epert\)-preserving if for every \( x \in \mathcal{X} \) and \( y \in \mathcal{Y} \),
\[
\Pr\Bigl[\mQ\bigl(x,\mP(x,y)\bigr) \ge \mQ(x,y)\Bigr] \ge \epert.
\]
\end{definition}

\begin{definition} [\cite{wits}, Definition 8]
\label{def:stationary-distribution}
    Let \( G = (V, E) \) be a weighted directed graph, and \( \vec{P} \) be the transition matrix of \( G \). We say that \( \vec{\pi} \in \mathbb{R}^n \) is a stationary distribution for \( \vec{P} \) if: $ \vec{P}^\top \cdot \vec{\pi} = \vec{\pi} $.
\end{definition}

\begin{definition}
\label{def:irreducibility}
    A weighted directed graph $ G = (\mathcal{V}, \mathcal{E}) $ is \textbf{irreducible} if for any pair of vertices $ u,v \in \mathcal{V}$, there exists a directed path from $ u $ to $ v $ with non-zero weight. In other words, there exists some $ t \geq 1$ such that $ \vec{P}^{t}(i,j) > 0 $. 
\end{definition}

\begin{definition}
\label{def:aperiodic}
    A weighted directed graph $ G = (\mathcal{V}, \mathcal{E}) $ is \textbf{aperiodic} if the greatest common divisor of the lengths of all directed cycles in $ G $ is 1.
\end{definition}

Let us now formally define the (hierarchically ordered) graph representations of $ \mP $ based on a prompt $ x \in \mathcal{X}$ and the quality threshold $ q \in [0,1]$.

\begin{definition}[\cite{wits}, Definition 7]
\label{def:graph_representation}
Fix an arbitrary prompt \( x \in \mathcal{X} \) and consider the graph \( \mG_{x} = (\vertex, \edge) \) whose vertex set is the output space of \( \mM \) (i.e., \( \vertex = \mathcal{Y} \)) and whose edge set \( \edge \) consists of all pairs \((y, y')\) such that
\[
\Pr\bigl[y' = \mP(x,y)\bigr] > 0.
\]

We assign weights \( w : \edge \to [0,1] \) to the edges by defining
\[
w(y, y') = \Pr\bigl[y' = \mP(x,y)\bigr].
\]
\end{definition}

Note that while the vertices of the graph are determined by the prompt \( x \in \mathcal{X} \) and the watermarking model \( \mM \), the edges and their weights are determined solely by \( \mP \). Let us now incorporate quality into the graph representation. Let $ \mG_{x}^{\geq q} $ be the subgraph of $ \mG_{x} $ given by

\[
\Vq = \{ y \in \mathcal{Y} \mid \mQ(x, y) \geq q \},
\]

\[
\Eq = \{ (y,y^{\prime}) \in \mathcal{Y} \times \mathcal{Y} \mid \mQ(x, y) \geq q, \mQ(x, y^{\prime}) \geq q,  \Pr\bigl[y' = \mP(x,y)\bigr] > 0\},
\]

Notice that we can carry the same weight assignment to this subgraph. Iteratively applying $ \mP $ on this graph and rejecting low-quality mutations produces a random walk where

\[
\vec{P}_{(y, y')} = \Pr\bigl[y' = \mP(x, y)\bigr].
\]

Before presenting the WITS impossibility result, we formally define watermarking schemes and related notions.

\begin{definition}[\cite{wits}, Definition 3]
\label{def:secret-key-wm-schemes}

Let $\mathcal{M} = \{\mM_i: X \to Y\}$ be a class of generative models with key space $K$. A secret-key watermarking scheme for $\mathcal{M}$ consists of two efficient algorithms:
\begin{itemize}
    \item \textbf{Watermark}$(\mM)$: A randomized algorithm that, given a model $\mM\in \mathcal{M}$, outputs a secret key $k\in K$ and a corresponding watermarked model $\mM_k: X \to Y$.
    \item \textbf{Detect}$_{k}(x,y)$: A deterministic algorithm that, given a secret key $k\in K$, a prompt $x\in X$, and an output $y\in Y$, returns a bit $b\in\{0,1\}$ indicating whether the watermark is present ($b=1$) or absent ($b=0$).
\end{itemize}
\end{definition}

We now define the false-positive rate $ \epsilon_{\text{pos}} $ of a watermarking scheme. Notice that if we sample $ y \xleftarrow{\$} \vec{\pi}$, we expect that $ y $ is unwatermarked with probability $ \epsilon_{\text{pos}} $. Since for every good watermarking scheme we need $ \epsilon_{\text{pos}} $ to be very small, this implies that outputs sampled from $ \vec{\pi} $ will be unwatermarked with high probability, i.e. with probability approximately $ 1 - \epsilon_{\text{pos}}$.

\begin{definition} [\cite{wits}, Definition 4]
\label{def:false-positive}
    A watermarking scheme \( \Pi \) has a false positive \( \epsilon_{\text{pos}} \)-rate if, for every model \( M \in \{ \mM_i : X \to Y \} \), for every prompt \( x \in X \), and for every output \( y \in Y \),

\[
\Pr[\text{Detect}_k(x, y) = 1] \leq \epsilon_{\text{pos}}.
\]

\end{definition}

Let us now define what it means for an adversary to break a watermarking scheme $ \Pi $. Notice that one might consider weaker notions of breaking a watermarking scheme, but they will be implied by the WITS result.

\begin{definition} [\cite{wits}, Definition 5]
\label{def:break-wm-scheme}
    
Let \( \Pi = (\text{Watermark}, \text{Detect}) \) be a watermarking scheme for a class of generative models \( \mathcal{M} = \{ \mM_i : X \to Y \} \) with associated quality function \( \mQ : X \times Y \to [0,1] \). 
We say that an adversary \( A \) \(\epsilon\)-breaks \( \Pi \) if for every \( M \in \mathcal{M} \), for every prompt \( x \in X \), we have:
\[ 
\Pr\big[\text{Detect}_k(x, y') = 0 \;\text{and}\; \mQ(x, y') \ge \mQ(x, y)
  : y \xleftarrow{\$} \mM_k(x),\; y' \xleftarrow{\$} A(x, y)\big] 
\;\ge\; \epsilon 
\]
where the probability is taken over \( (k, \mM_k) \) output by \( \text{Watermark}(M) \) and the random coins of \( A \).
\end{definition}

We now introduce an additional technical definition that enables the attacker to trade off between output quality and attack success probability. Let $v\in [0,100]$ denote the desired quality percentile. In other words, the attacker aims to produce an unwatermarked output whose quality falls within the top $v$-th percentile among all responses generated by $\mM$ on a given prompt $x$. To formalize this, define the set of achievable quality scores as
\[
\mQ_{M,x} = \Big\{ q : \Pr [ \mQ(x, \mM_k(x)) = q : (k, \mM_k) \xleftarrow{\$} \text{Watermark}(M) ] > 0 \Big\} \]
and let $q_{\mM,x}$ denote the $v$-th percentile of $\mQ_{M,x}$. We then define the overall minimum quality threshold as
\[
q_{\min} = \min_{\mM \in \mathcal{M},\, x \in \mathcal{X}} \left\{ q_{\mM,x} \right\}.
\]
We now state the WITS impossibility result.

\begin{theorem}[\cite{wits}, Theorem 6]
\label{theorem:wits_main_theorem}
Let $\Pi = (\text{Watermark}, \text{Detect})$ be a watermarking scheme for a class of generative models 
$\mathcal{M} = \{\mM_i : \mathcal{X} \to \mathcal{Y}\}$ with an associated quality function $\mQ : \mathcal{X} \times \mathcal{Y} \to [0,1]$. 
Let $\mP : \mathcal{X} \times \mathcal{Y} \to \mathcal{Y}$ be a perturbation oracle (defined over the same prompt space $\mathcal{X}$ and output space $\mathcal{Y}$ as the class $\mathcal{M}$) with the same associated quality function $\mQ : \mathcal{X} \times \mathcal{Y} \to [0,1]$ as $\Pi$. For every non-watermarked model $ \mM \in \mathcal{M}$, for every prompt $ x \in \mathcal{X}$, for every quality $ q \in q_{\min}$, let $ \vec{\pi}_{x,q}$ be the unique stationary distribution of the transition matrix $\vec{P}_{x,q}$ of $G^{\geq q}_x$. Let $ n_{x,q} = \lvert \mathcal{V}^{\geq q}_x \rvert $,  $ \pi_{\min}^{(x,q)} = \min \{\vec{\pi}_{x,q} (1), \dots, \vec{\pi}_{x,q} (n_{x,q})\} $ and $ g $ be the second largest eigenvalue of $\vec{P}_{x,q}$ in terms of absolute value. Let $ t_{\text{err}} > 0$ be a tunable parameter. Let $t_{x,q}$ be the $\edist$-mixing time of $\vec{P}_{x,q}$, defined as follows:

\[
t_{x,q} = \omega \left( 
\frac{1}{1 - g } 
\cdot \log \left( \frac{1}{\pi_{\min}^{(x,q)} \cdot \edist} \right) 
\right)
\]

Assume the following holds:
\begin{enumerate}
    \item The watermarking scheme $\Pi$ has a false positive $\epsilon_{\text{pos}}$-rate;
    \item The perturbation oracle $\mP$ is $\epert$-preserving;
    \item For every non-watermarked model $\mM \in \mathcal{M}$, for every prompt $x \in \mathcal{X}$, for every quality $q \in [q_{\min},1]$, 
    the $q$-quality $x$-prompt graph representation $\mG^{\geq q}_x$ of $\mP$ is irreducible and aperiodic.
\end{enumerate}

Then, there exists an oracle-aided universal adversary $A^{\mP(\cdot, \cdot),\mQ(\cdot, \cdot)}$ that $\epsilon$-breaks $\Pi$ by submitting at most $t$ queries to $\mP$ where

\[
\epsilon = \left( 1 - \frac{v}{100} \right) (1 - \epsilon_{\text{pos}}) (1 - \edist) \left( 1 - \sum_{k=0}^{t - t_{\text{err}} - 1} \binom{t}{k} (\epert)^k (1 - \epert)^{t-k} \right),
\]

and

\[
t = \max_{x \in \mathcal{X}, q \in [q_{\min},1]} \{t_{x,q}\} + t_{\text{err}}.
\]

\end{theorem}

By carefully tuning the parameter $t_{\text{err}}$ and running the attack long enough so that $\edist$ becomes negligibly small, the adversary can achieve a success probability close to
\[
\left(1 - \frac{v}{100}\right)(1 - \epsilon_{\text{pos}}).
\]
For example, targeting the median quality output (i.e., setting $v = 50$) restricts the adversary's success probability to roughly half of the maximum achievable rate.
\section{Appendix: Evaluation Setup}

\subsection{Watermark Details}
\label{appendix:watermark_details}

For KGW and SIR, we use the implementations contained within the \texttt{MarkLLM} package\footnote{\url{https://github.com/THU-BPM/MarkLLM}} \cite{markllm} with their default configurations. For Adaptive, we used the author's implementation\footnote{\url{https://github.com/yepengliu/adaptive-text-watermark}} and due to initially poor results, experimented heavily with different configurations to find one that best balanced initial quality and detectability for \texttt{Llama-3.1-70B-Instruct}. The three tunable parameters we explored were \texttt{alpha}, which thresholds the amount of token entropy required to watermark it; \texttt{delta}, which controls the strength of boosting for watermarked tokens; and \texttt{delta\_0}, which is the strength for watermarking the first $M=50$ tokens, which are always watermarked. Our analysis lead us to use \texttt{alpha} = 2.0, \texttt{delta} = 1.5, \texttt{delta\_0} = 1.0. Despite this extensive search, we still encountered intermittent issues with controlling for generation length. Since all texts were capped at a maximum of 1024 tokens due to fixed input sizes for various embedding models, some Adaptive responses were truncated mid-sentence, contributing to their unusually high number of grammatical errors as seen in Table~\ref{table:watermark_dataset}. 

\begin{figure}[htbp]
    \centering
    \includegraphics[width=0.85\textwidth]{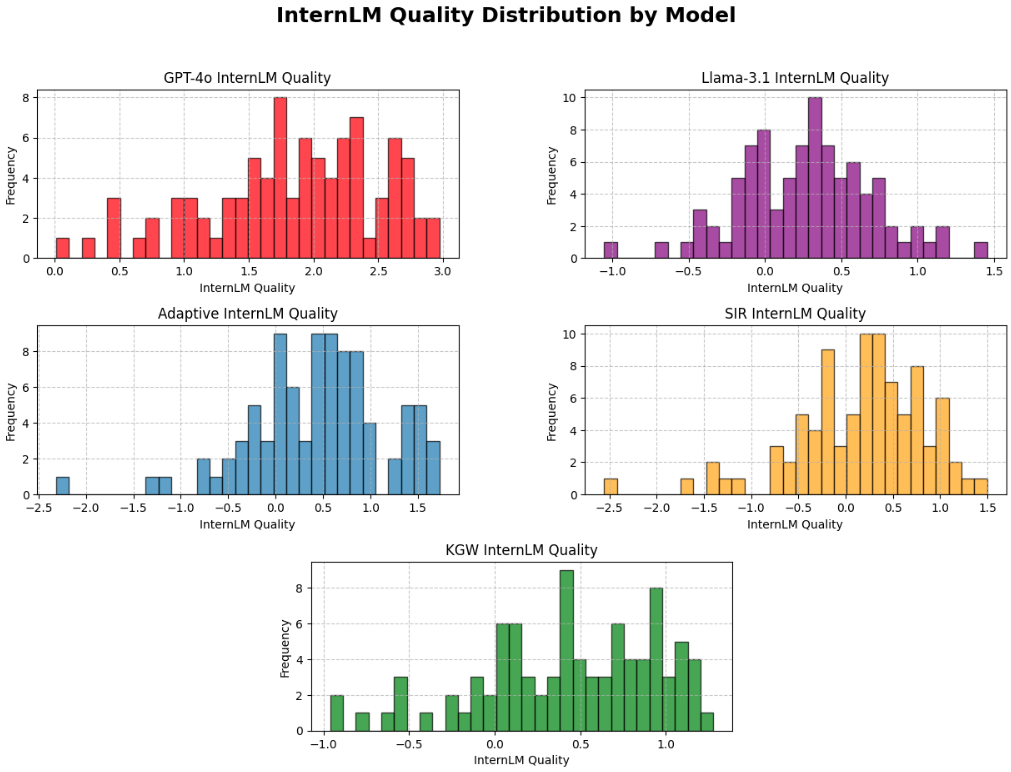}
    \caption{InternLM Quality Distribution by Watermarking Scheme $\mW$}
    \label{figure:internlm_quality}
\end{figure}

\begin{figure}[htbp]
    \centering
    \includegraphics[width=0.85\textwidth]{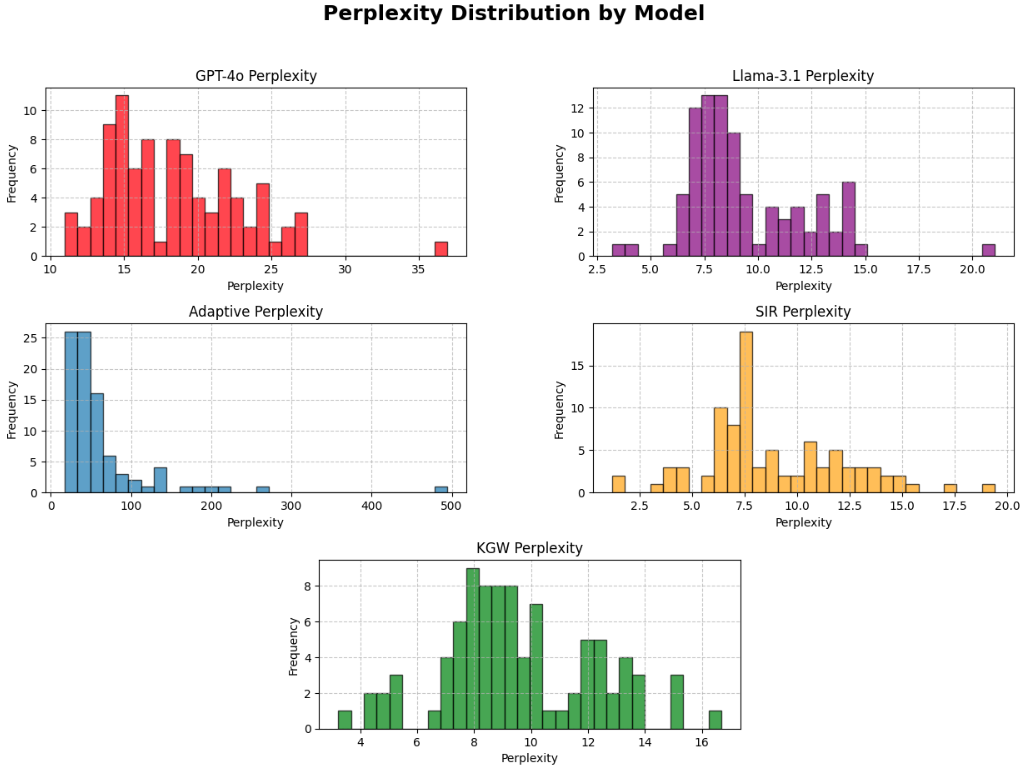}
    \caption{Perplexity Distribution by Watermarking Scheme $\mW$}
    \label{figure:perplexity}
\end{figure}

\begin{figure}[htbp]
    \centering
    \includegraphics[width=0.85\textwidth]{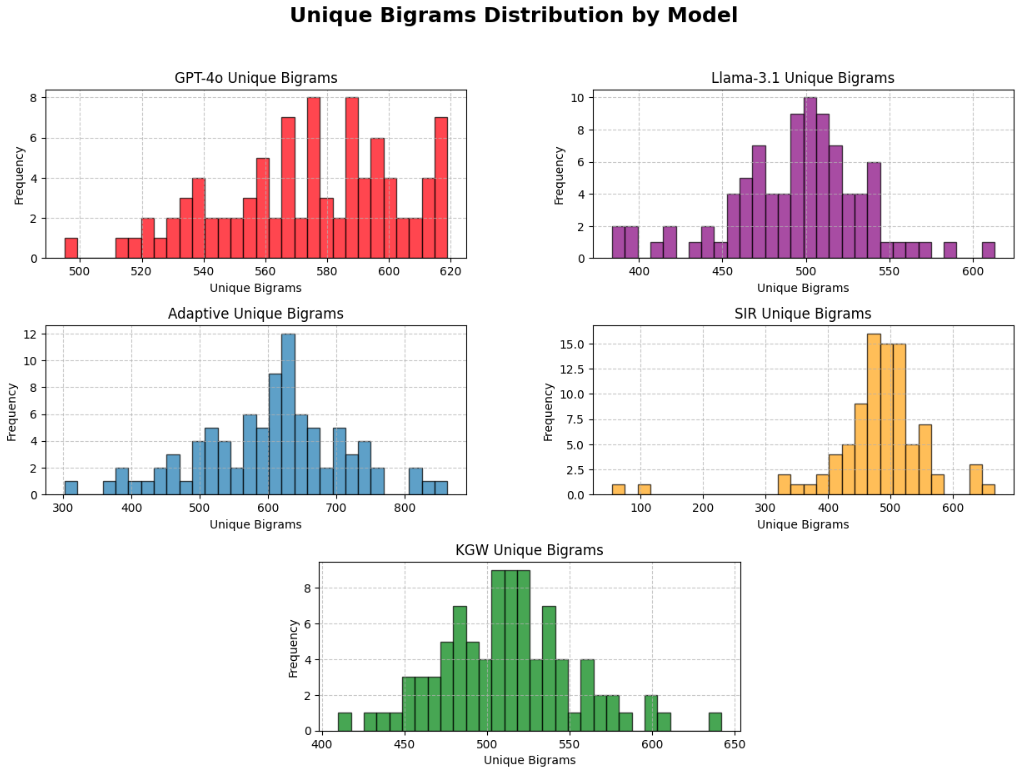}
    \caption{Unique Bigrams Distribution by Watermarking Scheme $\mW$}
    \label{figure:diversity}
\end{figure}

\begin{figure}[htbp]
    \centering
    \includegraphics[width=0.85\textwidth]{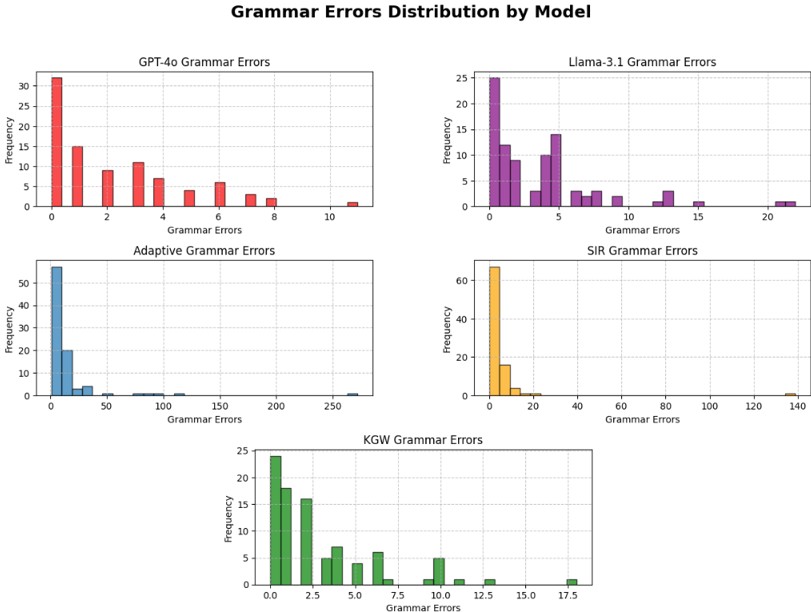}
    \caption{Grammar Errors Distribution by Watermarking Scheme $\mW$}
    \label{figure:grammar_errors}
\end{figure}

By plotting the distributions for each quality metric in Figures~\ref{figure:internlm_quality} through \ref{figure:grammar_errors}, we noticed that Adaptive and SIR were vulnerable to producing highly distorted text with numerous quality issues. For example, a single Adaptive generation contained over 250 grammatical issues, largely due to inexplicable letter case alterations (e.g. ``Over ThE nexT FEw dAYs, maggie partICIpaTed EnThusiasticALly I-n All ThE acTivities OffeRed aT WIllOW crEeEk...''). We did not regenerate bad responses because the distortions were a natural consequence of the watermarking algorithm itself, and regenerating them would obscure an important challenge to their real-world use. If the algorithm produces highly distorted text in some cases, then an attack is actually more likely to \emph{repair} quality, rather than merely preserving it. At least some cases in our study fit this profile and the attack should be fairly credited even if it generally does not work for texts that start at a higher standard of quality. 

% Nevertheless, Table~\ref{table:outlier_free_watermark_dataset} shows the automated quality scores after excluding responses with grammar errors beyond $\pm1.5$ times the interquartile range, which may better reflect the underlying effects of each watermark on quality and diversity. 

% \input{tables/outlier_free_watermark_dataset}

% InternLM Quality Figure
\medskip

% Perplexity Figure
\medskip

% Unique Bigrams Figure
\medskip

% Grammar Errors Figure

% Maintain the existing appendix flow without starting a new one
\clearpage

\subsection{Entropy-Controlled Prompt Dataset}
\label{appendix:entropy_prompts}

To systematically evaluate the impact of response entropy on watermark robustness, we curated a dataset featuring increasingly specific prompts across three domains: \textbf{creative writing}, \textbf{education}, and \textbf{journalism}. For each domain, we start with a broad, high-entropy prompt and progressively add constraints to reduce entropy. Below, we illustrate this progression with representative prompts at entropy level 1 (least constrained), 5, and 10 (most constrained).

\begin{table*}[!htbp]
\centering
\begin{tabular}{cl}
\toprule
\textbf{Entropy Level} & \textbf{Prompt} \\
\midrule
\multicolumn{2}{l}{\textbf{Creative Writing}} \\
1 & Write a 500-word story. \\
5 & Write a 500-word story about Evan, an American tourist, who falls for Emilie, a barista, \\
  & during a spring festival in Paris. \\
10 & Write a 500-word story about Evan, an American tourist, who falls for Emilie, a barista, \\
   & during a spring festival in Paris. They bond over their love for Claude Monet's \\
   & 'Impression, Sunrise' and the Hotel de Sully's architecture, leading to walks along the \\
   & Seine. Their connection deepens amid shared laughter and explorations of Le Marais. \\
   & As the festival lights dance on the river, Evan shares his feelings with Emilie under \\
   & the starlit sky, promising to cherish the moments they've shared. \\

\midrule
\multicolumn{2}{l}{\textbf{Education}} \\
1 & Write a 500-word essay about the importance of space exploration. \\
5 & Write a 500-word essay about the importance of space exploration, its role in advancing \\
  & human knowledge, and its potential to address global challenges like climate change \\
  & and resource scarcity, with a focus on technologies developed for space missions. \\
10 & Write a 500-word essay about the importance of space exploration, its role in advancing \\
   & human knowledge, and its potential to address global challenges like climate change \\
   & and resource scarcity, with a focus on technologies developed for space missions, \\
   & their applications on Earth, the possibility of colonizing other planets like Mars, \\
   & the ethical considerations of interplanetary exploration, and the cultural significance \\
   & of humanity becoming an interstellar species. \\

\midrule
\multicolumn{2}{l}{\textbf{Journalism}} \\
1 & Write a 500-word news article. \\
5 & Write a 500-word news article about a global climate summit where world leaders are \\
  & discussing strategies to combat climate change, with a focus on renewable energy \\
  & investments and carbon reduction targets, highlighting a groundbreaking agreement \\
  & between the US and China. \\
10 & Write a 500-word news article about a global climate summit where world leaders are \\
   & discussing strategies to combat climate change, with a focus on renewable energy \\
   & investments and carbon reduction targets, highlighting a groundbreaking agreement \\
   & between the US and China, featuring perspectives from small island nations affected \\
   & by rising sea levels, addressing protests outside the summit calling for stronger \\
   & climate justice measures, covering a controversial speech by a major oil industry \\
   & representative, analyzing the summit's key outcomes and challenges, and placing it in \\
   & the broader context of international efforts to achieve net-zero emissions by 2050. \\

\bottomrule
\end{tabular}
\caption{Representative entropy-controlled prompts across three domains: creative writing, education, and journalism. Entropy increases by adding specificity, progressively constraining the response space.}
\label{table:entropy_prompts}
\end{table*}

\newpage
\subsection{Dataset Statistics}
\label{appendix:dataset_statistics}

% \usepackage{graphicx}
\begin{table*}[!htbp]
\centering
\begin{tabular}{lrrrrr}
\toprule
 & \multicolumn{2}{c}{\textbf{Unwatermarked}}               & \multicolumn{3}{c}{\textbf{Watermarkered}}       \\
 & \multicolumn{1}{l}{\textbf{GPT-4o}} & \textbf{Llama-3.1} & \textbf{Adaptive} & \textbf{SIR} & \textbf{KGW} \\
 \midrule
Mean Watermarked Score ($\mu_{\text{w}}$)                      & --     & --     & 99.27  & 0.28   & 5.32   \\
Mean Unwatermarked Score ($\mu_{\text{uw}}$)             & --     & --     & 49.43  & 0.08   & -0.83  \\
Unwatermarked Standard Deviation  ($\sigma_{\text{uw}}$) & --     & --     & 3.37   & 0.07   & 1.05   \\
Breakpoint (Score $\leq \mu_{\text{uw}} + 2\sigma_{\text{uw}}$)    & --     & --     & 56.16  & 0.21   & 1.27   \\
\midrule
Quality Score                       & 1.85   & 0.27   & 0.45   & 0.16   & 0.43   \\
Perplexity                                   & 18.39  & 9.38   & 63.32  & 8.87   & 9.56   \\
Grammar Errors                               & 2.20   & 3.69   & 16.21  & 4.24   & 2.86   \\
Unique Bigrams Diversity                     & 574.06 & 494.59 & 603.64 & 479.90 & 512.44 \\
\midrule
Mean Word Count                              & 637.93 & 633.00 & 646.62 & 675.47 & 666.84 \\
Generation Time (s)                          & 15.24  & 274.27 & 671.75 & 335.12 & 292.24 \\
Detection Time (s)                           & --     & --     & 240.77 & 5.78   & 0.15   \\
\bottomrule
\end{tabular}
\caption{Summary statistics for unwatermarked and watermarked text across different watermarking schemes, highlighting detection scores, automatic quality metrics, and runtime statistics.}
\label{table:watermark_dataset}
\end{table*}

\newpage

\subsection{Perturbation Oracle Details}
\label{appendix:perturbation_oracles}

The perturbation oracles $\mP$ define the mechanism by which adversarial modifications are applied to watermarked text. These oracles generate perturbations of varying granularity, from token-level edits to full-document paraphrasing, enabling a systematic analysis of their impact on watermark robustness. Since prior work, including \citet{wits}, has not accounted for how different perturbation strategies affect attack success, we explore a diverse set of perturbation oracles to quantify their relative effectiveness.

\begin{itemize}[align=left, labelwidth=0.5cm, labelsep=0.15cm, leftmargin=0.75cm, noitemsep, nolistsep]  
    \item \textbf{\texttt{WordMutator}}: Randomly replaces individual tokens by masking and filling them using \texttt{RoBERTa} \cite{roberta}.
    \item \textbf{\texttt{EntropyWordMutator}}: Similar to \texttt{WordMutator}, but uses \texttt{GPT-Neo-2.7B} \cite{gpt-neo} to target high-entropy tokens for replacement as they are most likely to carry watermark signals.
    \item \textbf{\texttt{SpanMutator}}: Randomly masks six contiguous tokens at a time and fills them using \texttt{T5-XL} \cite{t5}. This is the only text-based perturbation oracle used in the WITS attack \cite{wits}.
    \item \textbf{\texttt{SentenceMutator}}: Randomly selects a sentence and paraphrases it creatively using \texttt{Llama-3.1-8B} \cite{llama3}, introducing higher-level semantic shifts.
    \item \textbf{\texttt{DocumentMutator}}: Uses the \texttt{DIPPER} paraphrase model \cite{dipper} to paraphrase multiple sections of the document simultaneously.
    \item \textbf{\texttt{Document1StepMutator}}: Re-generates the entire document from scratch using \texttt{Llama-3.1-8B}, producing the most extreme form of perturbation while preserving meaning, quality, and formatting.
    \item \textbf{\texttt{Document2StepMutator}}: Performs a two-step transformation, first selecting a random sentence and paraphrasing it creatively with \texttt{Llama-3.1-8B}, then performs a global consistency editing to ensure that the remaining text is consistent with the edited sentence.
\end{itemize}

These perturbation oracles serve two key purposes in our study: (1) they enable us to analyze how the size of the perturbation affects movement within the perturbation graph $\mG$; and (2) they allow us to determine whether specific perturbation oracles are more effective at breaking watermarks. By systematically evaluating these oracles, we aim to establish whether certain perturbation strategies inherently favor watermark removal and whether prior work has underestimated their impact on attack success. 

To ensure sufficient opportunity for watermark removal, we allow a large number of perturbation steps, proportional to the average number of words edited per step. For example, WordMutator is given 1000 steps, while DocumentMutator is given 100. Additionally, we note that each perturbation oracle was carefully calibrated to balance subtle modifications with sufficient impact on watermark signals, ensuring reproducibility by fixing random elements such as token selection and sampling temperature. Table~\ref{table:mutator_statistics} reveals a clear trade-off: while fine-grained oracles tend to preserve fluency, coarse-grained methods introduce larger variations -- a difference that is partly mitigated by the consistency editing in the \texttt{Document2StepMutator}.

\begin{table*}[!htbp]
    \centering
    \resizebox{\textwidth}{!}{%
    \begin{tabular}{lrrrrrrrr}
        \toprule
            \textbf{$\mP$} & 
            \textbf{Steps} & 
            \textbf{Edits} & 
            \textbf{PPL $\downarrow$} & 
            \textbf{Gram Err $\downarrow$} & 
            \textbf{Approval $\uparrow$} & 
            \textbf{Blocked $\downarrow$} & 
            \textbf{QScore $\uparrow$} & 
            \textbf{Time (s) $\downarrow$} \\
        \midrule
        \texttt{Word}          & 1000              & 1.8                   & 40.4                & 10.2                    & 0.80                   & 0                   & -0.0688                         & \textbf{0.10}                        \\
        \texttt{EntropyWord}   & 1000              & 1.1                   & 16.8                & 9.2                     & \textbf{0.82}          & 0                   & -0.0949                         & 0.28                                 \\
        \texttt{Span}          & 250               & 8.7                   & 27.2                & 7.8                     & 0.67                   & 0                   & -0.0746                         & 0.77                                 \\
        \texttt{Sentence}      & 150               & 31.3                  & 21.0                & 4.6                     & 0.74                   & 0                   & 0.2065                          & 0.94                                 \\
        \texttt{Document}      & 100               & 216.0                 & 11.2                & 9.1                     & 0.36                   & 0.12                & -0.3151                         & 213.12                             \\
        \texttt{Document1Step} & 100               & 138.2                 & \textbf{10.6}       & \textbf{2.1}            & 0.42                   & 0.12                & 0.0536                          &  29.61                              \\
        \texttt{Document2Step} & 100               & 105.2                 & 14.9                & 5.2                     & 0.54                   & 0.03                & \textbf{0.2089}                 &    34.78                                \\ 
        \bottomrule
    \end{tabular}%
    }
    \caption{Performance metrics for each perturbation oracle. The columns report the number of attack steps, average edits per step, average text perplexity (PPL), average number of grammar errors, average $\mQ$ approval rate, average rate at which $\mQ$ blocks every perturbation for a given prompt, average \texttt{InternLM} quality score (QScore), and average runtime per perturbation step in seconds. Emboldened values denote the best performance per metric.}
    \label{table:mutator_statistics}
\end{table*}

\section{Appendix: Distinguisher Prompt}
\label{appendix:distinguisher_prompt}

Designing a distinguisher requires careful consideration, as failures can be difficult to attribute to either implementation flaws or inherent limitations due to sufficient mixing. To address this challenge, we dedicated significant effort to prompt engineering, developing 12 different distinguishers and systematically benchmarking their performance. Somewhat unexpectedly, one of the simplest approaches proved to be the most effective, and we describe it below.

An important implementation detail is ensuring that our distinguisher does not pass tests due to random chance (as there are only 2 options it must choose between). Each test consists, on average, of 10 periodically sampled texts throughout the random walk. The test is considered a failure if even one sample fails, ensuring that the distinguisher can accurately deduce the origin throughout the entire attack. Additionally, we find that when the distinguisher fails, it almost always selects response A as the origin. We believe this to be due to positional bias. To combat this, every sample is distinguished as a best-of-2, where the second trial swaps response A and response B. The sample is considered a failure if either trial fails, minimizing the chance that the distinguisher succeeds by guessing. We remark that in every instance where the distinguisher failed, the best-of-2 was a tie, and zero samples (out of 57650) were incorrectly distinguished in both trials. This suggests the model was sometimes unable to overcome positional bias, but was never directly wrong. In these cases, we defer the final verdict to a stronger model, ultimately resulting in 100\% accuracy.

The following Python code defines the main method of \texttt{SimpleDistinguisher}, implemented with \texttt{guidance} \cite{guidance} to support constrained decoding. This function is designed to analyze response similarity while leveraging a predefined persona. The persona guides the AI model in determining whether a given perturbed response (\texttt{P}) originated from one of two original responses (\texttt{A} or \texttt{B}). 

\begin{lstlisting}[caption=Distinguisher Prompt Template, style=pythonstyle]
persona = \
"""
You are an expert in analyzing the similarity of responses.
You specialize in detecting whether one response has been derived from another by carefully analyzing the content and structure of the text.
Your expertise in linguistic analysis allows you to distinguish which responses are the most closely related.
Your goal is to provide a clear, concise, and accurate assessment of the provided instructions.
"""

@guidance
def distinguish(lm, persona, **kwargs):
    if persona:
        with system():
            lm += f"{persona}"
    with user():
        lm += f"""
        ### Task Description:
        1. Read the following 3 responses. They are labelled "A", "B", and "P".
        2. "A" and "B" are original responses, and "P" was generated by perturbing one of the original responses.
        3. Identify which of the original responses the perturbed story was derived from.
        
        Original response A:
        {kwargs["A"]}

        Original response B:
        {kwargs["B"]}

        Perturbed response P:
        {kwargs["P"]}
        """
    with assistant():
        lm += f"""\
        I believe the perturbed response P was derived from original response {select(["A", "B"], name="choice")}.
        """
    return lm
\end{lstlisting} 

\subsection{A Challenging Distinguisher Example}

As an example of the data used with the prompt above, we present an example that was particularly challenging for our distinguishers. After 108 sentence-level perturbations, \texttt{GPT-4o} was unable to accurately distinguish the origin. With around 30 seconds of reasoning, \texttt{o3-mini-high} correctly distinguished the origin, but had to correct itself while reasoning. We remark that the final section on NASA's Artemis program makes distinguishing this example trivial for humans, suggesting that our distinguishers are significantly weaker than humans. The perturbed text, along with the two original responses, are provided below with some key phrases in bold. 

\vspace{5mm}
\textbf{Perturbed Text (GPT-4o Failed to Distinguish)}  
\begin{quote}
Venturing into space is a groundbreaking endeavor that unlocks a multitude of benefits, extending far beyond the realms of scientific discovery and territorial growth. Space exploration, frequently overlooked, is a catalyst for scientific progress, driving the development of pioneering technologies and addressing humanity's most pressing challenges directly, making it a pursuit of paramount importance that warrants greater acknowledgment and support. This essay examines the importance of space exploration, its potential to broaden our understanding, and its ability to contribute to resolving critical global challenges like environmental decay and resource exhaustion. Understanding the cosmos is vital, as it allows us to grasp the intricate mechanisms \textbf{governing the universe and our place within it}, ultimately expanding our comprehension of reality itself. Exploring the vastness of the universe reveals a profound comprehension of the fundamental laws that shape reality, the origin of life, and the intricate chronology of cosmic evolution that has spanned eons of time.  Delving deeper into our environment not only quenches our innate desire for knowledge but also empowers us to make more informed choices about the planet's destiny, thereby shaping our relationship with the world that surrounds us. The pursuit of space exploration has far-reaching consequences, resulting in numerous groundbreaking discoveries that cumulatively contribute to a significant improvement in global well-being, manifesting in a multitude of tangible advantages. The rapid evolution of technology, encompassing satellite communication, GPS, and medical imaging, has significantly influenced our daily routines, work, and relationships, transforming the way we interact and live our lives. Advances in technology have not only bridged the world but have also led to better health outcomes worldwide, significantly impacting our daily lives and perceptions. Beyond its contributions to science and technology, space exploration provides a \textbf{distinctive vantage point for understanding the Earth and its interconnected systems}. Viewing our planet from space offers a comprehensive understanding of the interconnectedness of Earth's atmospheric, oceanic, and terrestrial systems, showcasing a cohesive entity that surpasses its individual components in complexity and unity. Understanding the effects of human actions on the environment is crucial for tackling pressing global issues, such as climate change, which is becoming more apparent with each passing day. Satellite imagery has been instrumental in tracking climate shifts, monitoring the growth of our oceans, and forecasting extreme weather events, all of which are crucial for comprehending the intricate dynamics of our planet's ever-changing environment. Creating a resilient and lasting future demands a sophisticated understanding of climate change's multifaceted impacts and the strategic application of targeted solutions to minimize its effects with accuracy and efficiency. One of the most significant benefits of space exploration is its potential to alleviate the consequences of dwindling resources on Earth. As the global population grows, the need for fundamental resources like food, water, and power is escalating, necessitating a significant boost in production to meet the rising demand. Exploring space can reveal previously unknown sources of essential materials, including minerals and water, which can be exploited to satisfy the growing need for these vital resources. \textbf{By 2024, NASA's Artemis mission seeks to create a self-sustaining lunar community}, ultimately paving the way for a permanent human settlement on the Moon that enables ongoing exploration and habitation. This expedition may reveal new helium-3 deposits, a scarce isotope that could serve as a fuel for nuclear fusion reactions, offering a potential source of power. Ultimately, space exploration is crucial for expanding our understanding of the universe, fostering groundbreaking technological advancements, and tackling pressing global issues such as environmental degradation and resource depletion. Venturing into the unknown reaches of space not only expands our comprehension of the cosmos, but also paves the way for groundbreaking discoveries that, in turn, significantly influence our daily lives and overall well-being. Satellite imaging provides a distinct viewpoint, allowing us to comprehend the intricacies of the obstacles we encounter and tackle them with greater success. As we venture deeper into the cosmos, we may stumble upon innovative answers to humanity's most critical challenges, thereby ensuring a thriving and sustainable world for future generations to inherit.
\end{quote}

\textbf{Original Response A}  
\begin{quote}
Space exploration is an essential endeavor for humanity, offering numerous benefits that extend beyond the realm of scientific discovery. While some may view it as a luxury or a frivolous pursuit, the reality is that space exploration plays a vital role in advancing human knowledge, driving technological innovation, and addressing some of the most pressing global challenges we face today. In this essay, we will explore the importance of space exploration, its contributions to human understanding, and its potential to help mitigate global crises like climate change and resource scarcity. One of the primary reasons space exploration is essential is that it allows us to expand our \textbf{understanding of the universe and our place within it}. By exploring the cosmos, we gain valuable insights into the fundamental laws of physics, the origins of life, and the evolution of the universe. This knowledge not only satisfies our innate curiosity but also informs our understanding of the world around us, enabling us to make more informed decisions about our planet and its resources. Furthermore, space exploration drives technological innovation, which has numerous spin-off benefits for society. The development of new technologies, such as satellite communications, GPS, and medical imaging, has transformed the way we live, work, and interact with one another. These innovations have also improved our daily lives, from enabling global connectivity to enhancing healthcare outcomes. In addition to its scientific and technological benefits, space exploration also offers a \textbf{unique perspective on the Earth and its systems}. By studying the Earth from space, we gain a deeper understanding of the complex interactions between our atmosphere, oceans, and landmasses. This knowledge is essential for addressing global challenges like climate change, where the impacts of human activity on the planet are becoming increasingly evident. Space-based observations have played a critical role in monitoring climate change, tracking sea-level rise, and predicting weather patterns. These insights are essential for developing effective strategies to mitigate the effects of climate change and ensure a sustainable future. Another significant benefit of space exploration is its potential to address resource scarcity. As the global population continues to grow, the demand for resources like food, water, and energy is increasing. Space exploration offers a unique opportunity to identify new sources of resources, such as minerals and water, that can help meet this demand. For example, \textbf{NASA's Artemis program aims to return humans to the Moon by 2024}, with the long-term goal of establishing a sustainable presence on the lunar surface. This mission could potentially unlock new resources, such as helium-3, a rare isotope that could be used as fuel for nuclear fusion. In conclusion, space exploration is essential for advancing human knowledge, driving technological innovation, and addressing global challenges like climate change and resource scarcity. By exploring the cosmos, we gain valuable insights into the universe and our place within it, while also developing new technologies that improve our daily lives. Furthermore, space-based observations offer a unique perspective on the Earth and its systems, enabling us to better understand and address the complex challenges we face. As we continue to explore the vast expanse of space, we may uncover new solutions to some of humanity's most pressing problems, ensuring a sustainable and prosperous future for generations to come.
\end{quote}

\textbf{Original Response B}  
\begin{quote}
Space exploration is an essential endeavor for humanity, offering numerous benefits that extend beyond the realm of scientific discovery. It plays a vital role in advancing human knowledge, driving technological innovations, and addressing pressing global challenges such as climate change and resource scarcity. In this essay, we will explore the significance of space exploration and its potential to shape the future of our planet and beyond. The pursuit of space exploration is often viewed as a costly and ambitious endeavor, but it is essential to recognize the significant contributions it makes to our \textbf{understanding of the universe and the world we inhabit}. By venturing into space, we gain insights into the fundamental laws of physics, the origins of life, and the evolution of the cosmos. These discoveries not only expand our scientific knowledge but also inspire new generations of scientists, engineers, and innovators. Furthermore, space exploration has led to numerous technological innovations that have transformed various aspects of our daily lives. From the development of GPS and telecommunications to medical imaging and weather forecasting, the spin-off benefits of space exploration have been substantial. These innovations have improved the quality of life for millions of people around the world and have also generated significant economic benefits. In addition to its scientific and technological benefits, space exploration also offers a \textbf{unique opportunity to address pressing global challenges}. For instance, the study of Earth from space provides critical insights into the health of our planet and the impacts of climate change. Satellite imaging and remote sensing technologies have enabled scientists to monitor deforestation, track ocean currents, and detect changes in global temperature patterns. This information is essential for developing effective strategies to mitigate the effects of climate change and promote sustainable development. Another significant benefit of space exploration is its potential to provide new resources and opportunities for economic growth. As the world's population continues to grow, the demand for resources such as food, water, and energy will increase. Space exploration offers a way to address this challenge by accessing new sources of resources, such as asteroid mining and lunar helium-3 extraction. These resources could provide a clean and sustainable source of energy, reducing our reliance on fossil fuels and mitigating the impacts of climate change. Finally, space exploration offers a unique opportunity for international cooperation and diplomacy. In an era marked by increasing global tensions and conflict, space exploration provides a shared goal that can bring nations together. Collaborative efforts such as the International Space Station and the \textbf{Artemis program} have demonstrated the potential for space exploration to foster global cooperation and understanding. In conclusion, space exploration is essential for advancing human knowledge, driving technological innovations, and addressing pressing global challenges. Its significance extends beyond the realm of scientific discovery, offering numerous benefits that have the potential to shape the future of our planet and beyond. As we continue to explore the vastness of space, we must recognize the importance of investing in this endeavor and working together to address the challenges that lie ahead. By doing so, we can ensure that the benefits of space exploration are shared by all and that the next generation of scientists, engineers, and innovators is inspired to reach for the stars.
\end{quote}

\noindent

\section{Extended Distinguisher Study}
\label{appendix:extended_distinguisher_study}

In addition to the main \textbf{RQ1} result, we designed an even more challenging evaluation setting to test whether sufficient mixing could obscure the lineage of perturbed texts. Specifically, we focus on the strongest $\mP$, \texttt{SentenceMutator}, as it previously demonstrated the highest capacity to evade detection by \texttt{Llama-3.1-70B}. To amplify its effect, we increase the perturbation budget from 150 to 500 steps, allowing the random walk significantly more opportunities to approach the stationary distribution.

Additionally, we constrain the attack to texts generated from the lowest-entropy prompts, ensuring that candidate parent texts are highly similar to one another. This combination of (1) the strongest perturbation oracle, (2) an extended attack budget, and (3) a highly confounded candidate pool creates the most difficult setting for lineage attribution. If mixing is truly effective under these conditions, we would expect distinguishability to approach random chance.

We find that although the task was more challenging, with more failures on average, \texttt{o3-mini-high} still had no issues in distinguishing the origin in each test.

\begin{table*}[!htbp]
\centering
\begin{tabular}{lrrrrr}
\toprule
\textbf{$\mP$ Oracle}           & \textbf{Steps}                & \textbf{Tests}                & \textbf{\texttt{Llama-3.1-70B}} & \textbf{\texttt{GPT-4o}} & \textbf{\texttt{o3-mini-high}}   \\
\midrule
\texttt{Sentence}                & 500                          & 54                           & 13                               & 3                        & 0                \\
\midrule
\textbf{Cumulative Distinguished (\%)} & \multicolumn{1}{l}{\textbf{}} & \multicolumn{1}{l}{\textbf{}} & \textbf{75.9}                  & \textbf{94.4}           & \textbf{100}     \\
\bottomrule
\end{tabular}
\caption{Summary of failed distinguisher tests on the most challenging settings. Classification is first performed by \texttt{Llama-3.1-70B}, followed by \texttt{GPT-4o} on its failures, then \texttt{o3-mini-high} on any remaining cases. The overall 100\% success rate indicates that the attacked texts never lose memory of their starting points, contradicting \textbf{KA1} and suggesting that a stationary distribution is not reached in practice.}
\label{table:distinguisher_long}
\end{table*}

\subsection{Breakdown by Domain and Entropy}
We find domain to be significant in distinguishability, but surprisingly, not entropy.

\begin{table*}[!htbp]
\centering
\begin{tabular}{lrr}
\toprule
\textbf{Domain}     & \textbf{Failed Distinguishes (Main)} & \textbf{Failed Distinguishes (Challenge)} \\
\midrule
Journalism          & 6/1458                               & 0/18                                      \\
Creative Writing    & 7/1560                               & 6/18                                      \\
Education           & 40/1537                              & 7/18                                      \\
\bottomrule
\end{tabular}
\caption{Domain distribution for tests which \texttt{Llama-3.1-70B} failed to distinguish.}
\label{table:distinguisher_domain}
\end{table*}
\begin{table*}[!htbp]
\centering
\begin{tabular}{rrr}
\toprule
\multicolumn{1}{l}{\textbf{Entropy}} & \textbf{Failed Distinguishes (Main)} & \textbf{Failed Distinguishes (Challenge)} \\
\midrule
1                                    & 7/462                                & N/A                                       \\
2                                    & 4/457                                & N/A                                       \\
3                                    & 8/468                                & N/A                                       \\
4                                    & 9/462                                & N/A                                       \\
5                                    & 3/450                                & N/A                                       \\
6                                    & 1/468                                & N/A                                       \\
7                                    & 6/450                                & N/A                                       \\
8                                    & 2/438                                & N/A                                       \\
9                                    & 5/456                                & N/A                                       \\
10                                   & 8/444                                & 13/54                                     \\
\bottomrule
\end{tabular}
\caption{Entropy distribution for tests which \texttt{Llama-3.1-70B} failed to distinguish.}
\label{table:distinguisher_entropy}
\end{table*}

\newpage
\section{Appendix: Quality Oracles}

\subsection{Oracle Details}
\label{appendix:oracles}

The quality oracles determine whether the perturbations introduced by various $\mP$ preserve the original text’s quality. Each oracle operates by querying an LLM with a prompt and some continuation text using different strategies to assess preservation of meaning, fluency, and coherence. The quality decision is based on whether the mutated text is judged to be as good as or better than the original. All oracle queries include the original prompt to provide context for evaluation.

We implement and evaluate eight distinct quality oracles using \texttt{guidance} \cite{guidance} to support constrained decoding for ranking, scoring, and preference based assessments. 

\begin{itemize}[align=left, labelwidth=0.5cm, labelsep=0.15cm, leftmargin=0.75cm, noitemsep, nolistsep]
    \item \textbf{\texttt{RankOracle}}: $\mQ$ is prompted to rank the two responses in terms of preference, and the order of texts is then reversed in a second query. If the mutated text is preferred in both cases, quality is considered preserved.
    \item \textbf{\texttt{SoloOracle}}: $\mQ$ is prompted twice, independently grading each text on a numerical scale. If the mutated text receives a score equal to or higher than the original, its quality is considered preserved.
    \item \textbf{\texttt{JointOracle}}: Similar to \texttt{SoloOracle}, but $\mQ$ assigns numerical scores to both texts in the same prompt. The order is flipped in a second query. Quality is preserved if the mutated text scores equal to or higher than the original in both cases.
    \item \textbf{\texttt{RelativeOracle}}: $\mQ$ is prompted to select the better response or declare a tie, repeating the query with the order reversed. Quality is preserved if the mutated text is chosen in both cases or a tie is declared.
    \item \textbf{\texttt{BinaryOracle}}: $\mQ$ is asked a direct yes/no question: ``Is the second text just as good or better than the original?'' If the response is ``yes'', quality is preserved.
    \item \textbf{\texttt{MutationOracle}}: Similar to \texttt{BinaryOracle}, but the prompt explicitly states that the second text is a modification of the original. The query is repeated with the order reversed. If both responses are ``yes'', quality is preserved.
    \item \textbf{\texttt{ExampleOracle}}: Similar to \texttt{BinaryOracle}, but includes an example (1-shot prompting) before presenting the actual texts. If the response is ``yes'', quality is preserved.
    \item \textbf{\texttt{DiffOracle}}: $\mQ$ is provided with the original text, mutated text, and a computed diff between them. It is asked whether these changes are acceptable. If the response is ``yes'', quality is preserved.
\end{itemize}

These oracles serve as key components in our evaluation framework, allowing us to systematically assess how to best approximate human judgments of quality. By incorporating multiple prompting strategies, we ensure robustness in our analysis of watermark perturbation effectiveness.

\subsection{Sandcastle Dataset Statistics}
\label{appendix:sandcastle_details}

Since absolute quality scoring is difficult for humans \cite{chatbotarena}, we formulated the annotation task as pairwise preference judgments with a tie option. Several coauthors, following standardized guidelines, compared perturbed texts to their originals, unaware of which was which. Table \ref{table:sandcastles_dataset} shows the class distribution, where we merged "Attacked Better" and "Tie" into a Quality Preserved (QP) category to support binary classification.

\begin{table}[!htbp]
\centering
\begin{tabular}{lrrrr}
\toprule
\textbf{}      & \multicolumn{2}{c}{\textbf{Quality Preserved}} & \multicolumn{1}{c}{\textbf{Quality Degraded}} & \multicolumn{1}{l}{}               \\
\textbf{Split} & \textbf{Attacked Better}       & \textbf{Tie}       & \textbf{Original Better}                      & \multicolumn{1}{l}{\textbf{Total}} \\
\midrule
Train          & 12                             & 238                & 306                                           & 556                                \\
Test           & 1                              & 103                & 135                                           & 239                                \\
\midrule
Total          & 13                             & 341                & 441                                           & 795                                \\
\bottomrule
\end{tabular}
\caption{Distribution of human quality assessments by split for the Sandcastles dataset. The table details counts for cases where attacked outputs were rated as "Attacked Better" or "Tie" (grouped under Quality Preserved (QP)) versus "Original Better", along with overall totals for both training and test sets.}
\label{table:sandcastles_dataset}
\end{table}

\newpage
\subsection{Full Oracle Results}

Table~\ref{table:oracles_full} provides a detailed comparison of quality oracles, including inference time, QP Precision, Overall F1, and RewardBench scores where available. Despite fine-tuning, no oracle fully aligns with human judgments, and high RewardBench scores do not guarantee strong performance in our setting. Proprietary models like \texttt{GPT-4o} with fine-tuning perform best but are impractical for large-scale attacks. Locally hosted models (\texttt{MutationOracle}, \texttt{DiffOracle}) offer a viable alternative but still misclassify degraded outputs. These results highlight the challenges of using LLM-based oracles for reliable watermark attack guidance.

\begin{table*}[!htbp]
\centering
\resizebox{\textwidth}{!}{%
\begin{tabular}{lllrrrr}
\toprule
\textbf{Oracle}     & \textbf{Model}                                                                                                                & \textbf{Type} & \textbf{Time (s)} & \textbf{QP Prec.} & \textbf{Overall F1} & \textbf{RB Score} \\
\midrule
SkyworkOracle       & \href{https://huggingface.co/Skywork/Skywork-Reward-Gemma-2-27B-v0.2}{Skywork-Reward-Gemma-2-27B-v0.2} \cite{skywork} & FLOAT         & 2.22              & 43.51                 & 26.39               & 94.3              \\
RankOracle          & \href{https://huggingface.co/meta-llama/Llama-3.1-70B-Instruct}{Llama-3.1-70B-Instruct} \cite{llama3}                 & BOOL          & 4.33              & 50.00                 & 37.09               & --                \\
SoloOracle          & \href{https://huggingface.co/meta-llama/Llama-3.1-70B-Instruct}{Llama-3.1-70B-Instruct} \cite{llama3}                 & INT           & 2.23              & 49.49                 & 39.86               & --                \\
JointOracle         & \href{https://huggingface.co/meta-llama/Llama-3.1-70B-Instruct}{Llama-3.1-70B-Instruct} \cite{llama3}                 & INT           & 3.62              & 53.85                 & 40.85               & --                \\
INFORMOracle        & \href{https://huggingface.co/infly/INF-ORM-Llama3.1-70B}{INF-ORM-Llama3.1-70B} \cite{inform}                          & FLOAT         & 5.81              & 65.63                 & 54.40               & 95.1              \\
QRMOracle           & \href{https://huggingface.co/nicolinho/QRM-Gemma-2-27B}{QRM-Gemma-2-27B} \cite{qrmoracle}                             & FLOAT         & 3.28              & 50.68                 & 56.98               & 94.4              \\
RelativeOracle      & \href{https://huggingface.co/meta-llama/Llama-3.1-70B-Instruct}{Llama-3.1-70B-Instruct} \cite{llama3}                 & CLASS         & 2.76              & 79.59                 & 63.07               & --                \\
ExampleOracle       & \href{https://huggingface.co/meta-llama/Llama-3.1-70B-Instruct}{Llama-3.1-70B-Instruct} \cite{llama3}                 & BOOL          & 1.33              & 79.59                 & 63.07               & --                \\
BinaryOracle        & \href{https://huggingface.co/meta-llama/Llama-3.1-70B-Instruct}{Llama-3.1-70B-Instruct} \cite{llama3}                 & BOOL          & 1.27              & 61.90                 & 63.82               & --                \\
ArmoRMOracle        & \href{https://huggingface.co/RLHFlow/ArmoRM-Llama3-8B-v0.1}{ArmoRM-Llama3-8B-v0.1} \cite{armorm}                      & FLOAT         & 0.33              & 65.71                 & 64.26               & 90.4              \\
OffsetBiasOracle    & \href{https://huggingface.co/NCSOFT/Llama-3-OffsetBias-RM-8B}{Llama-3-OffsetBias-RM-8B} \cite{offsetbias}             & FLOAT         & 0.32              & 62.22                 & 65.30               & 89.6              \\
Prometheus2Absolute & \href{https://huggingface.co/prometheus-eval/prometheus-8x7b-v2.0}{prometheus-8x7b-v2.0}  \cite{prometheus2}          & FLOAT         & 7.28              & 74.78                 & 66.73               & 74.5              \\
Prometheus2Relative & \href{https://huggingface.co/prometheus-eval/prometheus-8x7b-v2.0}{prometheus-8x7b-v2.0}  \cite{prometheus2}          & BOOL          & 7.36              & 74.78                 & 66.73               & 74.5              \\
Prometheus2Absolute & GPT-4o \cite{gpt4o}                                                                                                   & INT           & 7.93              & 76.70                 & 66.87               & --                \\
MutationOracle      & \href{https://huggingface.co/meta-llama/Llama-3.1-70B-Instruct}{Llama-3.1-70B-Instruct} \cite{llama3}                 & BOOL          & 2.74              & \textbf{84.62}        & 66.93               & --                \\
Prometheus2Relative & GPT-4o \cite{gpt4o}                                                                                                   & BOOL          & 7.73              & 77.23                 & 67.05               & --                \\
Prometheus2Relative & GPT-4-Turbo \cite{gpt4turbo}                                                                                          & BOOL          & 11.94             & 75.00                 & 67.27               & --                \\
Prometheus2Absolute & GPT-4-Turbo \cite{gpt4turbo}                                                                                          & INT           & 12.46             & 76.15                 & 67.55               & --                \\
InternLMOracle      & \href{https://huggingface.co/internlm/internlm2-20b-reward}{internlm2-20b-reward} \cite{internlm}                     & FLOAT         & 0.86              & 65.69                 & 69.84               & 90.6              \\
DiffOracle          & \href{https://huggingface.co/meta-llama/Llama-3.1-70B-Instruct}{Llama-3.1-70B-Instruct} \cite{llama3}                 & BOOL          & 1.83              & 71.74                 & 70.85               & --                \\
MutationOracle+FT   & \href{https://huggingface.co/meta-llama/Llama-3.1-70B-Instruct}{Llama-3.1-70B-Instruct} + Fine-tuning                 & BOOL          & 3.25              & 81.18                 & 71.83               & --                \\
DiffOracle+FT       & \href{https://huggingface.co/meta-llama/Llama-3.1-70B-Instruct}{Llama-3.1-70B-Instruct} + Fine-tuning                 & BOOL          & 1.80              & 69.07                 & 76.94               & --                \\
DiffOracle+FT       & GPT-4o \cite{gpt4o} + Fine-tuning                                                                                     & BOOL          & 0.46              & 75.51                 & 77.32               & --                \\
MutationOracle+FT   & GPT-4o \cite{gpt4o} + Fine-tuning                                                                                     & BOOL          & 0.84              & 74.51                 & \textbf{77.38}      & --                \\
\bottomrule
\end{tabular}%
}
\caption{Overview of oracle performance on our human-annotated test set. For each oracle we report average inference time, Quality-Preserved (QP) Precision, Overall F1, and RewardBench (RB) Score when available. Despite fine-tuning on human judgements, no oracle perfectly capture human quality assessments, and high RB Scores did not predict strong performance in our evaluation setting.}
\label{table:oracles_full}
\end{table*}

\newpage
\subsection{\texttt{InternLM} vs \texttt{DiffOracle}}
\label{appendix:i_vs_d}

We compared the proportion of cases where humans agreed that the quality of generated outputs was preserved. The results, summarized in Table~\ref{table:head2head}, show that \texttt{InternLMOracle} had a higher agreement rate (47.78\%) than \texttt{DiffOracle+FT} (40.0\%).

\begin{table}[!htbp]
\centering
\begin{tabular}{lrr}
\toprule
\textbf{Oracle} & \textbf{Agree QP} & \textbf{Disagree QP} \\
\midrule
DiffOracle      & 40.00             & 60.00                \\
InternLM        & 47.78             & 52.22                \\
\bottomrule
\end{tabular}
\caption{Comparison of human agreement rates on quality preservation (QP) percentages between DiffOracle and InternLM. InternLM shows a higher agreement rate, suggesting it aligns better with human judgments.}
\label{table:head2head}
\end{table}

To quantify the probability that \texttt{InternLMOracle} is genuinely the better oracle, we adopt a Bayesian approach, modeling the probability of agreement for each oracle as a Beta distribution:

\begin{equation*}
    \begin{aligned}
        p_A &\sim \text{Beta}(A+1, N_A-A+1), \\
        p_B &\sim \text{Beta}(B+1, N_B-B+1)
    \end{aligned}
\end{equation*}

where $A$ and $B$ are the counts of human agreement for \texttt{DiffOracle+FT} and \texttt{InternLMOracle}, respectively, and $N_A$ and $N_B$ are the total evaluations for each oracle.

Using a Monte Carlo simulation with 100,000 samples, we estimate:

\begin{equation*}
    P(p_B > p_A) \approx 85.08\%
\end{equation*}

indicating that \texttt{InternLMOracle} has an 85.08\% probability of being the better judge in preserving quality according to human evaluators. Given this high confidence, we justify the use of \texttt{InternLMOracle} as the preferred oracle for further evaluations\footnote{This surprising reversal of performance may be attributable to \texttt{DiffOracle+FT} managing too much noise in the changelog of edits when attacks exceed 20 steps (the maximum attack length present in the Sandcastles dataset).}.

\section{Appendix: Human Annotation Details}
\label{appendix:human_study_details}

Annotators were provided with the following instructions when reviewing:

\begin{itemize}
    \item Determine which is a better response to the prompt: text $A$, text $B$, or tie.
    \item Judge quality based on content, style, cohesion, and prompt relevance.
    \item Note: Formatting is not especially important for quality (e.g. paragraph breaks should be ignored). 
\end{itemize}

These guidelines ensured that evaluations focused on meaningful quality differences rather than superficial formatting artifacts.

\newpage
\section{Appendix: Extended Attack Results Analysis}
\label{appendix:full_attack_details}

Table~\ref{table:full_attack_results} provides a detailed breakdown of attack performance, including automated quality metrics, revealing several notable patterns. One interesting finding is that, in some cases, attacks appear to ``improve'' certain quality metrics, such as perplexity and grammar error rates. This effect is most pronounced for the Adaptive watermark, where the average perplexity and grammar errors decrease post-attack. However, this improvement is largely driven by a few low-quality outliers in the original watermarked dataset, rather than a systematic enhancement of text fluency. Despite these reductions in surface-level errors, the \texttt{InternLM} quality score consistently drops, indicating that attacks tend to reduce overall coherence and relevance, even when fluency-related metrics superficially improve.

Another trend is that unique bigram diversity (\(\mu_{d_t}\)) increases slightly in many cases, particularly for sentence- and document-level attacks. This suggests that perturbations introduce more varied word sequences, potentially disrupting structured patterns imposed by watermarking. However, this increase is relatively small, meaning that while attacks may inject lexical diversity, they do not necessarily enhance the text in a meaningful way. Instead, the most aggressive perturbation strategies—particularly sentence- and document-level attacks—cause the largest drops in the \texttt{InternLM} quality score, reinforcing the idea that these attacks are the most disruptive to text coherence. While they achieve the highest watermark removal rates, they also tend to introduce noticeable degradation, making the resulting text less natural and readable.

By contrast, perturbation strategies that fail to effectively break watermarks, such as word-level and entropy-based edits, also have minimal impact on quality metrics. This suggests that these finer-grained mutations are too minor to erase watermark signals while also being too weak to meaningfully degrade text fluency. More broadly, the average attack success rate remains relatively low even before enforcing quality constraints, with \(\text{ASR}_{\mathrm{fin}}\) at only 26.13\%. After accounting for quality degradation, this drops further to just 10.47\%, confirming that successfully removing watermarks without compromising text quality remains a substantial challenge for adversaries.

% \usepackage{lscape}
\begin{landscape}
\begin{table*}[!htbp]
\centering
\resizebox{\columnwidth}{!}{%
\begin{tabular}{llrrrrrrrrrrrrrrrrr}
\toprule
\textbf{Watermark} &
\textbf{$\mP$ Oracle} &
$\,\bm{\mu_{w_0}}$ &
$\,\bm{\mu_{w_t}}$ &
\textbf{BP} &
\textbf{$\text{ASR}_{\mathrm{min}}$} &
\textbf{$\text{ASR}_{\mathrm{fin}}$} &
\textbf{Reviewed} &
\textbf{QP} &
$\neg$\textbf{QP} &
\textbf{$\text{Q-ASR}_{\mathrm{fin}}$} &
$\,\bm{\mu_{q_0}}$ &
$\,\bm{\mu_{q_t}}$ &
$\,\bm{\mu_{p_0}}$ &
$\,\bm{\mu_{p_t}}$ &
$\,\bm{\mu_{g_0}}$ &
$\,\bm{\mu_{g_t}}$ &
$\,\bm{\mu_{d_0}}$ &
$\,\bm{\mu_{d_t}}$ \\
\midrule
\texttt{Adaptive} & \texttt{Word}          & 99.27 & 70.37 & 56.16 & 0.00  & 0.00  & 0  & 0  & 0  & 0.00  & 0.45 & 0.01  & 63.32 & 77.64 & 16.21 & 13.79 & 603.64 & 609.09 \\
\texttt{Adaptive} & \texttt{EntropyWord}   & 99.27 & 82.45 & 56.16 & 0.00  & 0.00  & 0  & 0  & 0  & 0.00  & 0.45 & -0.05 & 63.32 & 78.01 & 16.21 & 19.00 & 603.64 & 608.33 \\
\texttt{Adaptive} & \texttt{Span}          & 99.27 & 67.21 & 56.16 & 1.54  & 1.54  & 2  & 2  & 0  & 1.54  & 0.45 & 0.06  & 63.32 & 44.34 & 16.21 & 11.24 & 603.64 & 616.06 \\
\texttt{Adaptive} & \texttt{Sentence}      & 99.27 & 59.93 & 56.16 & 35.34 & 19.21 & 20 & 8  & 12 & 7.68  & 0.45 & 0.24  & 63.32 & 27.78 & 16.21 & 5.42  & 603.64 & 602.26 \\
\texttt{Adaptive} & \texttt{Document}      & 99.27 & 58.55 & 56.16 & 48.78 & 45.24 & 20 & 8  & 12 & 18.10 & 0.45 & 0.00  & 63.32 & 16.27 & 16.21 & 2.11  & 603.64 & 464.31 \\
\texttt{Adaptive} & \texttt{Document1Step} & 99.27 & 70.94 & 56.16 & 1.16  & 1.16  & 2  & 2  & 0  & 1.16  & 0.45 & 0.32  & 63.32 & 30.09 & 16.21 & 0.46  & 603.64 & 541.44 \\
\texttt{Adaptive} & \texttt{Document2Step} & 99.27 & 73.39 & 56.16 & 5.33  & 4.71  & 8  & 5  & 3  & 2.94  & 0.45 & 0.27  & 63.32 & 27.48 & 16.21 & 8.06  & 603.64 & 580.33 \\
\midrule
\texttt{KGW}      & \texttt{Word}          & 0.28  & 0.17  & 0.21  & 47.54 & 20.00 & 20 & 4  & 16 & 4.00  & 0.16 & -0.30 & 8.87  & 30.30 & 4.24  & 11.10 & 479.90 & 572.57 \\
\texttt{KGW}      & \texttt{EntropyWord}   & 0.28  & 0.22  & 0.21  & 3.45  & 0.56  & 1  & 0  & 1  & 0.00  & 0.16 & -0.31 & 8.87  & 20.15 & 4.24  & 10.14 & 479.90 & 535.14 \\
\texttt{KGW}      & \texttt{Span}          & 0.28  & 0.20  & 0.21  & 38.46 & 32.35 & 20 & 14 & 6  & 22.65 & 0.16 & -0.29 & 8.87  & 14.78 & 4.24  & 7.19  & 479.90 & 536.79 \\
\texttt{KGW}      & \texttt{Sentence}      & 0.28  & 0.14  & 0.21  & 89.47 & 56.52 & 20 & 7  & 13 & 19.78 & 0.16 & 0.02  & 8.87  & 16.93 & 4.24  & 4.68  & 479.90 & 658.71 \\
\texttt{KGW}      & \texttt{Document}      & 0.28  & 0.18  & 0.21  & 62.50 & 44.44 & 20 & 8  & 12 & 17.78 & 0.16 & -0.27 & 8.87  & 10.48 & 4.24  & 7.70  & 479.90 & 435.39 \\
\texttt{KGW}      & \texttt{Document1Step} & 0.28  & 0.27  & 0.21  & 12.66 & 8.54  & 14 & 7  & 7  & 4.27  & 0.16 & 0.07  & 8.87  & 9.47  & 4.24  & 0.31  & 479.90 & 482.51 \\
\texttt{KGW}      & \texttt{Document2Step} & 0.28  & 0.18  & 0.21  & 9.09  & 7.78  & 10 & 4  & 6  & 3.11  & 0.16 & -0.03 & 8.87  & 11.97 & 4.24  & 4.41  & 479.90 & 501.89 \\
\midrule
\texttt{SIR}      & \texttt{Word}          & 5.32  & 1.74  & 1.27  & 78.22 & 57.89 & 20 & 1  & 19 & 2.89  & 0.43 & 0.00  & 9.56  & 29.96 & 2.86  & 6.79  & 512.44 & 588.33 \\
\texttt{SIR}      & \texttt{EntropyWord}   & 5.32  & 3.30  & 1.27  & 39.68 & 27.54 & 20 & 0  & 20 & 0.00  & 0.43 & -0.03 & 9.56  & 20.20 & 2.86  & 1.44  & 512.44 & 554.38 \\
\texttt{SIR}      & \texttt{Span}          & 5.32  & 1.57  & 1.27  & 60.71 & 37.40 & 20 & 5  & 15 & 9.35  & 0.43 & 0.03  & 9.56  & 15.28 & 2.86  & 4.36  & 512.44 & 561.04 \\
\texttt{SIR}      & \texttt{Sentence}      & 5.32  & 0.52  & 1.27  & 87.65 & 74.71 & 20 & 13 & 7  & 48.56 & 0.43 & 0.30  & 9.56  & 17.97 & 2.86  & 2.93  & 512.44 & 597.44 \\
\texttt{SIR}      & \texttt{Document}      & 5.32  & 0.93  & 1.27  & 61.54 & 46.09 & 20 & 6  & 14 & 13.83 & 0.43 & 0.10  & 9.56  & 12.65 & 2.86  & 10.64 & 512.44 & 454.21 \\
\texttt{SIR}      & \texttt{Document1Step} & 5.32  & 2.54  & 1.27  & 14.04 & 14.04 & 12 & 11 & 1  & 12.87 & 0.43 & 0.23  & 9.56  & 11.74 & 2.86  & 0.26  & 512.44 & 491.45 \\
\texttt{SIR}      & \texttt{Document2Step} & 5.32  & 3.07  & 1.27  & 68.09 & 49.06 & 20 & 12 & 8  & 29.44 & 0.43 & 0.26  & 9.56  & 11.70 & 2.86  & 4.29  & 512.44 & 530.79 \\
\midrule
\textbf{Averages} &                        &       &       &       & 36.44 & 26.13 &    & 40.48 & 59.52   & 10.47 & 0.35 & 0.03  & 27.25 & 25.49 & 7.77  & 6.49  & 532.00 & 548.69 \\
\bottomrule
\end{tabular}%
}
\caption{Attack success rates (ASR) and autoamated quality scores across different perturbation strategies. Human review reveals an average of 59.52\% of successfully attacked texts have degraded quality. $\bm{\mu_{w_0}}$ represents the initial watermark score at step 0, while $\bm{\mu_{w_t}}$ represents the final watermark score after $t$ mutation steps.``min'' refers to the point where the watermark score is at its lowest during the attack while ``fin'' refers to score at the final step of the attack. "Reviewed" indicates the number of human-reviewed examples where the watermark was broken. \textbf{QP} and $\neg$\textbf{QP} represent the number of cases where human reviewers judged the attacked text as quality-preserving or degraded, respectively. \textbf{Q-ASR}$_{\mathrm{fin}}$ is the re-estimated attack success after controling for quality, calculated as \textbf{ASR}$_{\mathrm{fin}} \times  (\text{QP} / \text{Reviewed})$. The remaining quality columns show, for time step 0 and the final step $t$, \texttt{InternLM} quality score (q), perplexity (p), grammar errors (g), and unique bigram diversity (d). On average, quality degraded significantly while perplexity, grammar errors, and diversity improved.}
\label{table:full_attack_results}
\end{table*}
\end{landscape}

\subsection{Attack Success Rate vs.\ Detection Threshold}

\begin{figure}
    \centering
    \includegraphics[width=0.9\linewidth]{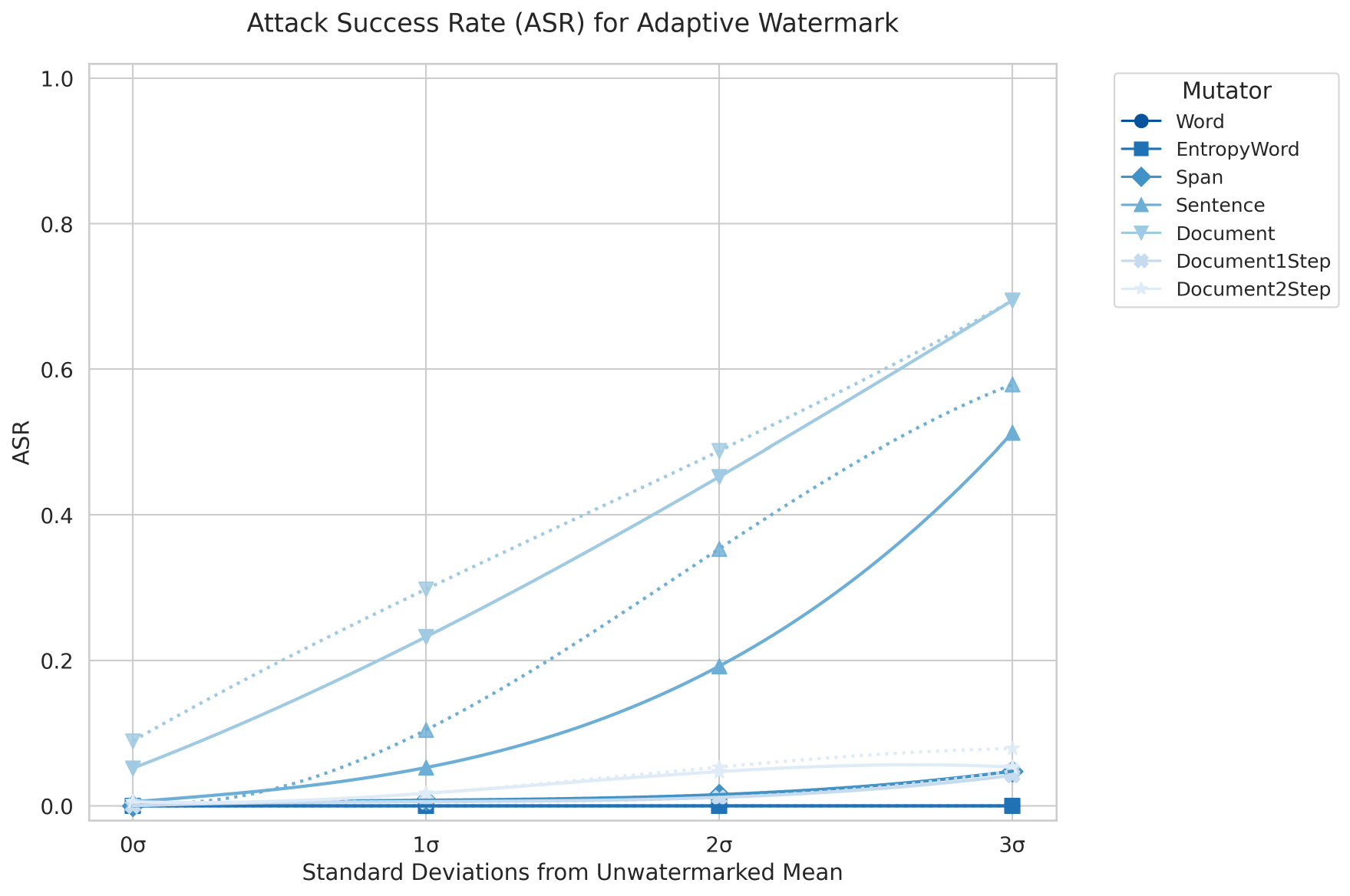}
    \caption{Attack success rate (ASR) vs. detection threshold for the Adaptive watermarking scheme. Each curve represents a different perturbation oracle, with thresholds measured in standard deviations above the unwatermarked mean.}
    \label{fig:asr_adaptive}
\end{figure}

\begin{figure}
    \centering
    \includegraphics[width=0.9\linewidth]{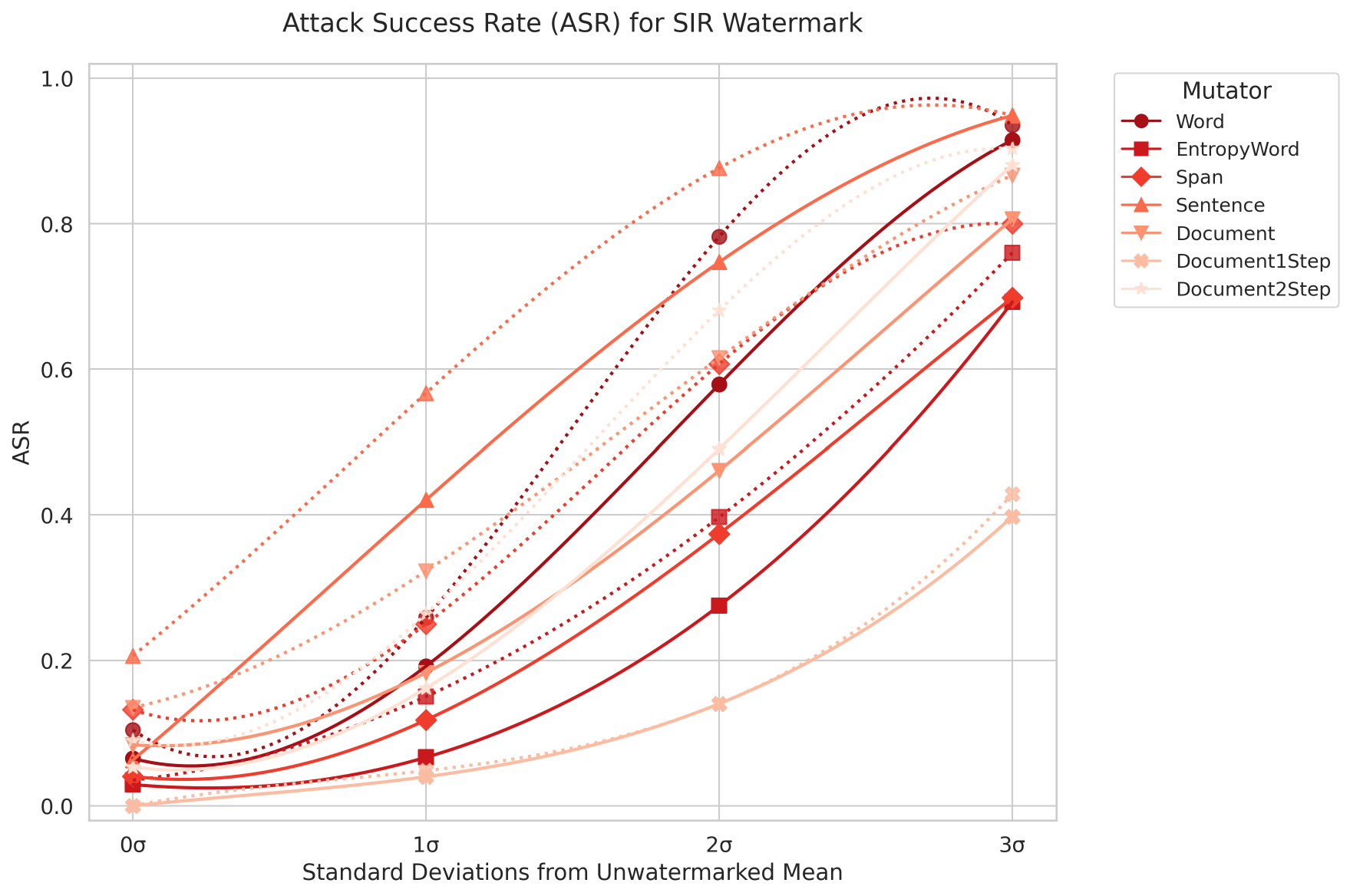}
    \caption{Attack success rate (ASR) vs. detection threshold for the SIR watermarking scheme. The plot shows the fraction of attacked texts falling below various thresholds (in standard deviations above the unwatermarked mean) for multiple perturbation oracles.}
    \label{fig:asr_sir}
\end{figure}

\begin{figure}
    \centering
    \includegraphics[width=0.9\linewidth]{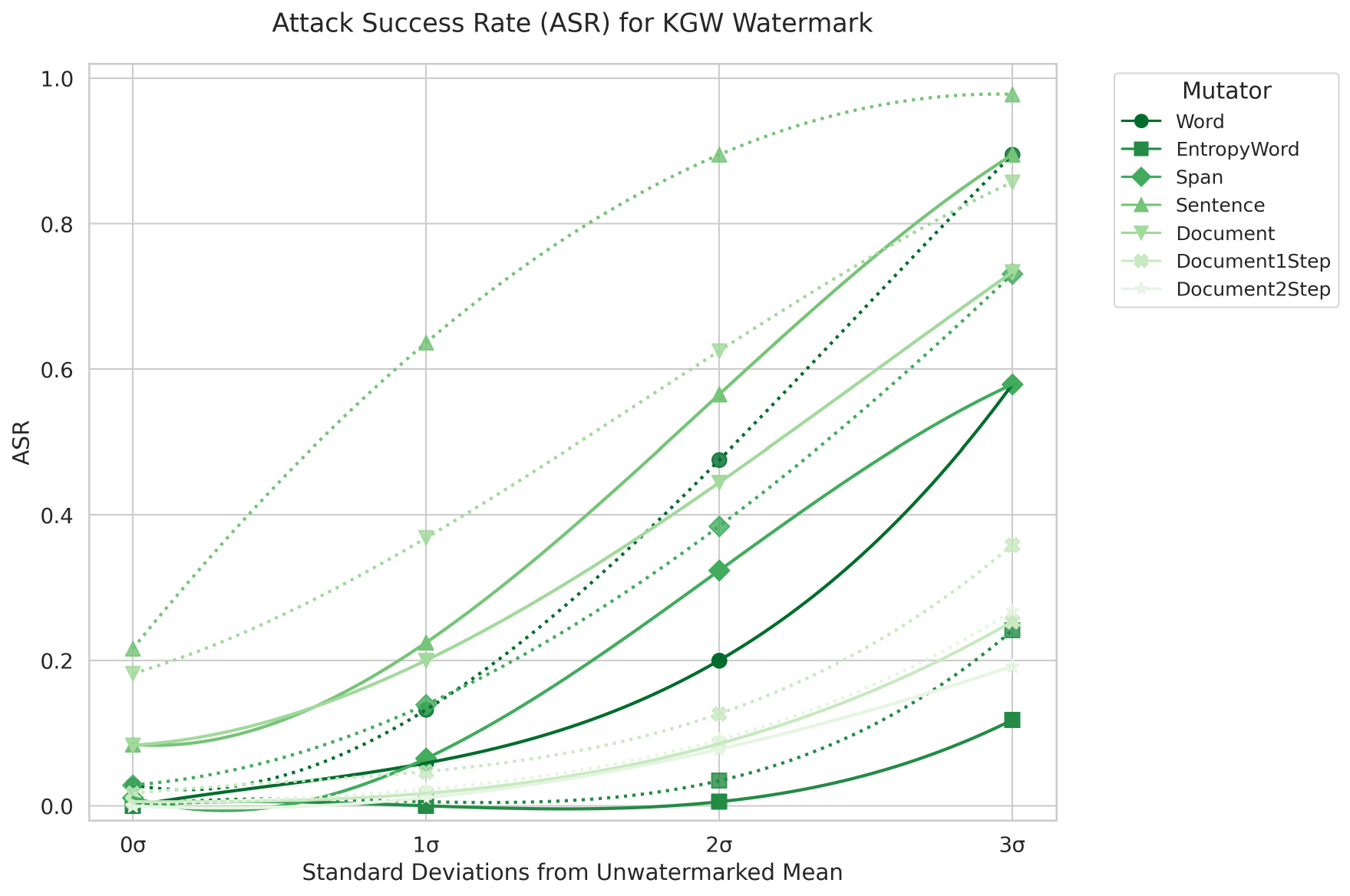}
    \caption{Attack success rate (ASR) vs. detection threshold for the KGW watermarking scheme. Different curves correspond to various perturbation oracles, with the detection threshold defined as standard deviations above the mean detection score of unwatermarked texts.}
    \label{fig:asr_kgw}
\end{figure}

Figures~\ref{fig:asr_kgw}, \ref{fig:asr_sir}, and \ref{fig:asr_adaptive} plot the final attack success rate (\textbf{ASR}) for each perturbation oracle under the \textbf{KGW}, \textbf{SIR}, and \textbf{Adaptive} watermarking schemes, respectively. The horizontal axis represents detection thresholds measured in standard deviations above the mean detection score for unwatermarked text (i.e., $0\sigma$, $1\sigma$, $2\sigma$, or $3\sigma$). A higher threshold allows more texts to be considered ``unwatermarked,'' so ASR generally increases as we move to the right. The vertical axis indicates the fraction of attacked texts that fall below each threshold once all permitted mutations have been applied.

Each curve corresponds to a specific mutator---\texttt{Word}, \texttt{EntropyWord}, \texttt{Span}, \texttt{Sentence}, \texttt{Document}, \texttt{Document1Step}, or \texttt{Document2Step}---with line style distinguishing $s_{min}$ (dotted) from $s_{fin}$ (solid). In general, token-level $\mP$ (\texttt{Word}, \texttt{EntropyWord}, \texttt{Span}) make smaller, more localized edits, while document-level $\mP$ (\texttt{Document}, \texttt{Document1Step}, \texttt{Document2Step}) can restructure larger portions of text. Comparing these curves reveals which $\mP$ achieve higher ASR for each watermarking scheme and how sensitive those results are to stricter or looser detection thresholds.

Overall, two main patterns emerge. First, as the detection threshold increases, more perturbed texts evade being flagged, causing the ASR curves to rise. Second, the extent of this rise varies across both watermarking schemes and $\mP$: some methods prove more effective at evading detection for KGW or SIR, whereas Adaptive typically shows lower ASR across thresholds. This aligns with our broader observation that larger, more context-aware edits (\texttt{Document}-based $\mP$) often outperform smaller, token-level edits, but still rarely achieve high success rates without risking noticeable quality degradation.

\newpage

\section{Appendix: What factors contributed to attack inefficiency?}
\label{appendix:attack_inefficiency}

The efficiency of the WITS attack against private watermarking schemes is hampered by two interrelated challenges. First, the attack relies on a random walk that must approach its stationary distribution, with the mixing time critically dependent on the second-largest eigenvalue, $g$, of the transition matrix $\vec{P}$. Not only is computing $g$ exactly infeasible, but even approximating it is extremely difficult. In practice, the size and complexity of $ \vec{P}$—which depends on factors such as the mutator, prompt, and quality barrier—make computing any information about $ \vec{P}$ computationally intractable. As a result, the attacker must rely on upper bounds for $g$ to estimate the mixing time, a strategy that introduces significant uncertainty into the overall attack duration. Notice that this isn't an issue for public watermarking schemes since the attacker can stop as soon as the watermark is removed.

Second, attempts to accelerate the mixing process—such as by increasing the step size of the perturbation oracle—risk degrading the quality of the text. As quality decreases, so does the success rate of mutations (i.e., the effective constant $\epert$ no longer holds), which in turn negates the benefits of improved mixing by requiring even more iterations to produce acceptable outputs.

In essence, there is a fundamental tension between reducing the mixing time to achieve attack efficiency and maintaining the quality of the attacked text. A more refined theoretical analysis that balances these competing factors is necessary to fully understand the capabilities of the WITS attack. We leave this compelling direction for future work.

\begin{figure}[htbp]
    \centering
    \includegraphics[width=0.95\textwidth]{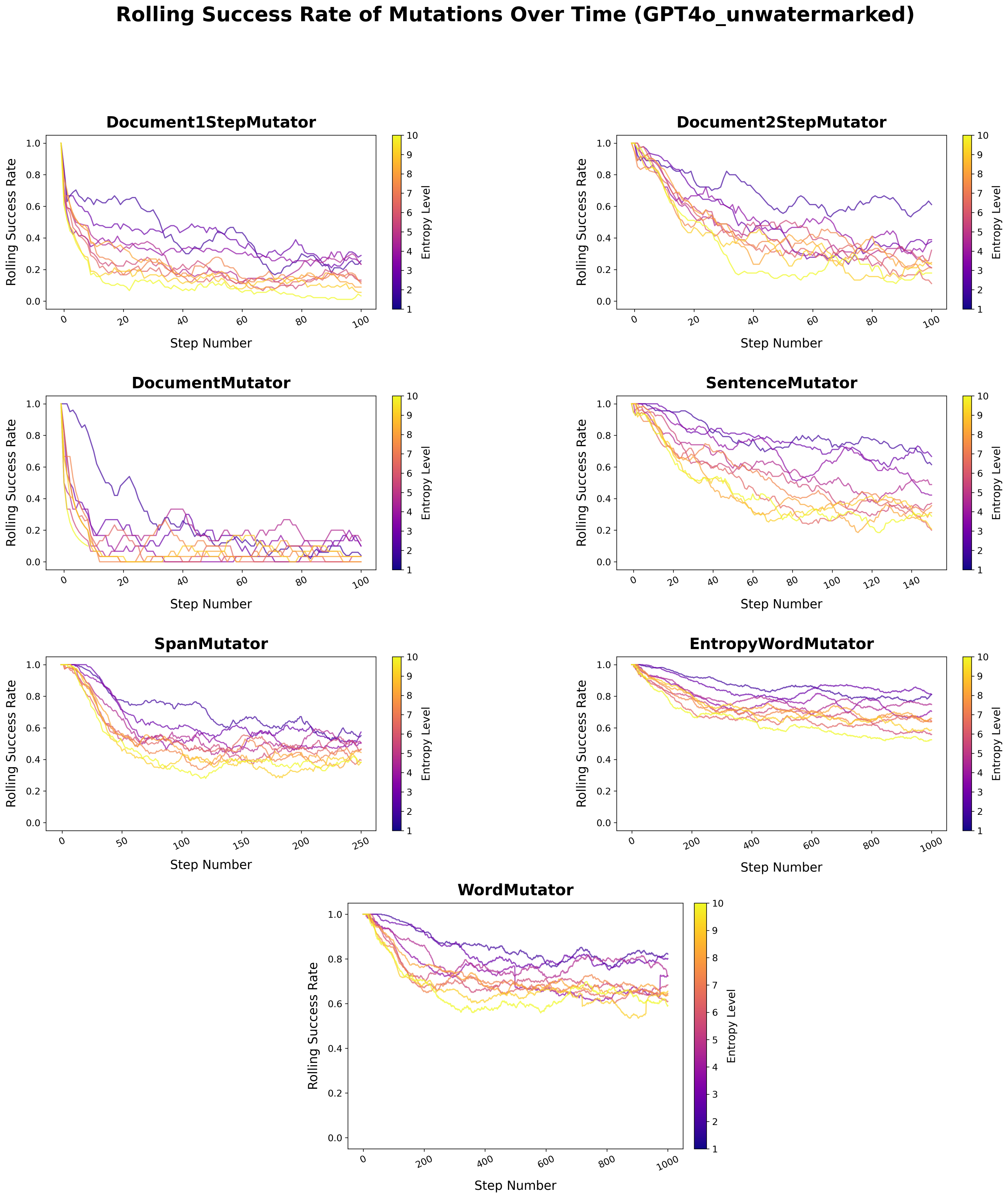}
    \caption{Rolling success rate for GPT-4o generations, which are unwatermarked.}
    \label{figure:rolling_success_GPT4o}
\end{figure}

\begin{figure}[htbp]
    \centering
    \includegraphics[width=0.95\textwidth]{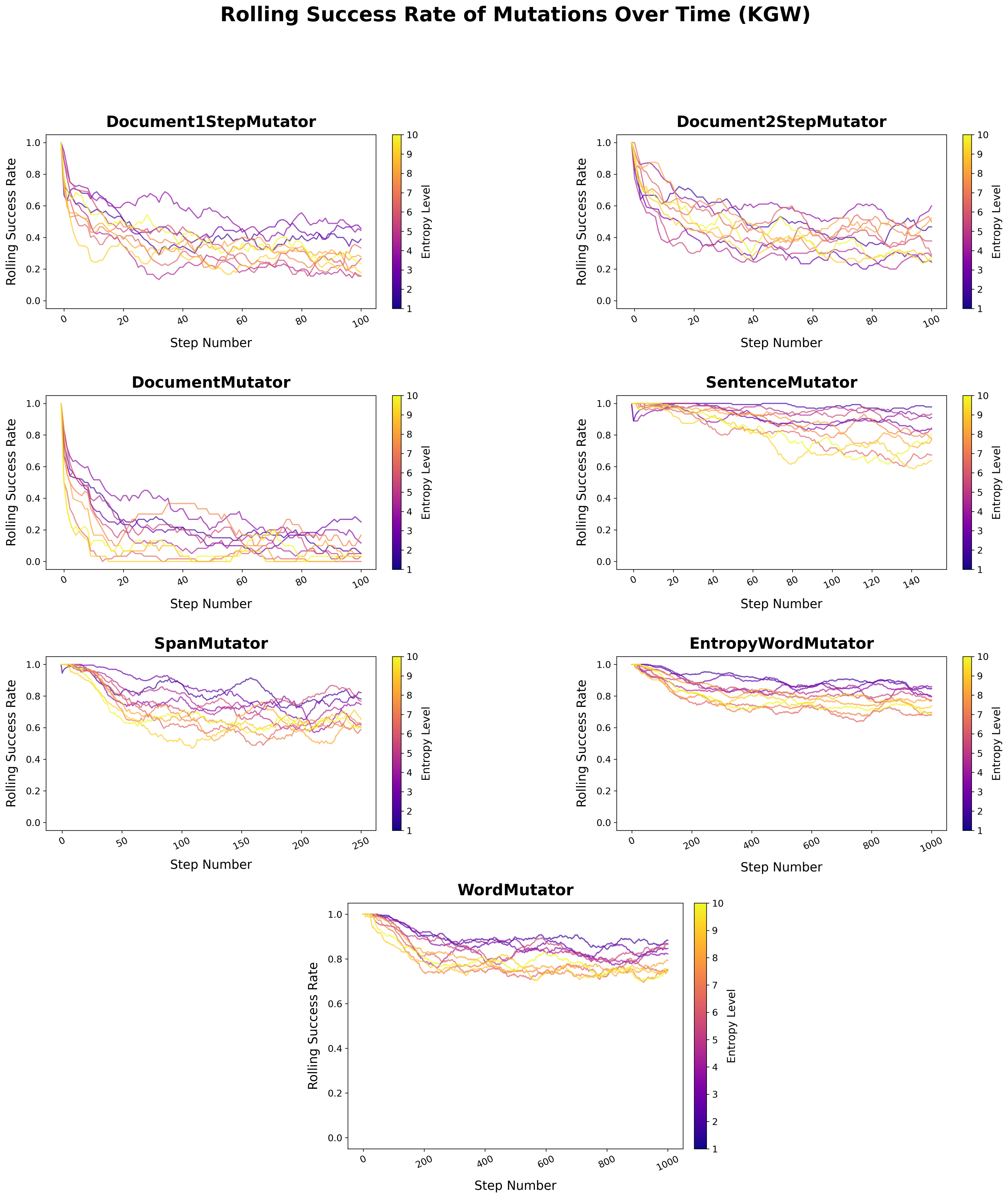}
    \caption{Rolling success rate for the KGW watermark.}
    \label{figure:rolling_success_KGW}
\end{figure}

Figures~\ref{figure:rolling_success_GPT4o}, \ref{figure:rolling_success_KGW}, \ref{figure:rolling_success_SIR}, and \ref{figure:rolling_success_Adaptive} below illustrate the rolling success rate of mutations across various watermarking schemes and mutator types, thereby supporting our first claim. In these computations, the window size is defined as one-tenth of the total number of mutator steps (e.g., for the Sentence Mutator, \(150/10 = 15\) steps).

Notably, $\mP$ characterized by larger step sizes exhibit lower success rates. Furthermore, the plots reveal a modest correlation between the mutation success rate and the entropy level: prompts with lower entropy tend to have reduced success rates. This phenomenon may be attributable to the fact that lower-entropy prompts are generally longer, thereby increasing the difficulty of generating a mutated response that maintains high quality. Consequently, any interpretation of this correlation should be approached with caution.

% Rolling Success GPT-4o (Unwatermarked) Figure

% Rolling Success KGW Figure

% Rolling Success SIR Figure
\begin{figure}[htbp]
    \centering
    \includegraphics[width=0.95\textwidth]{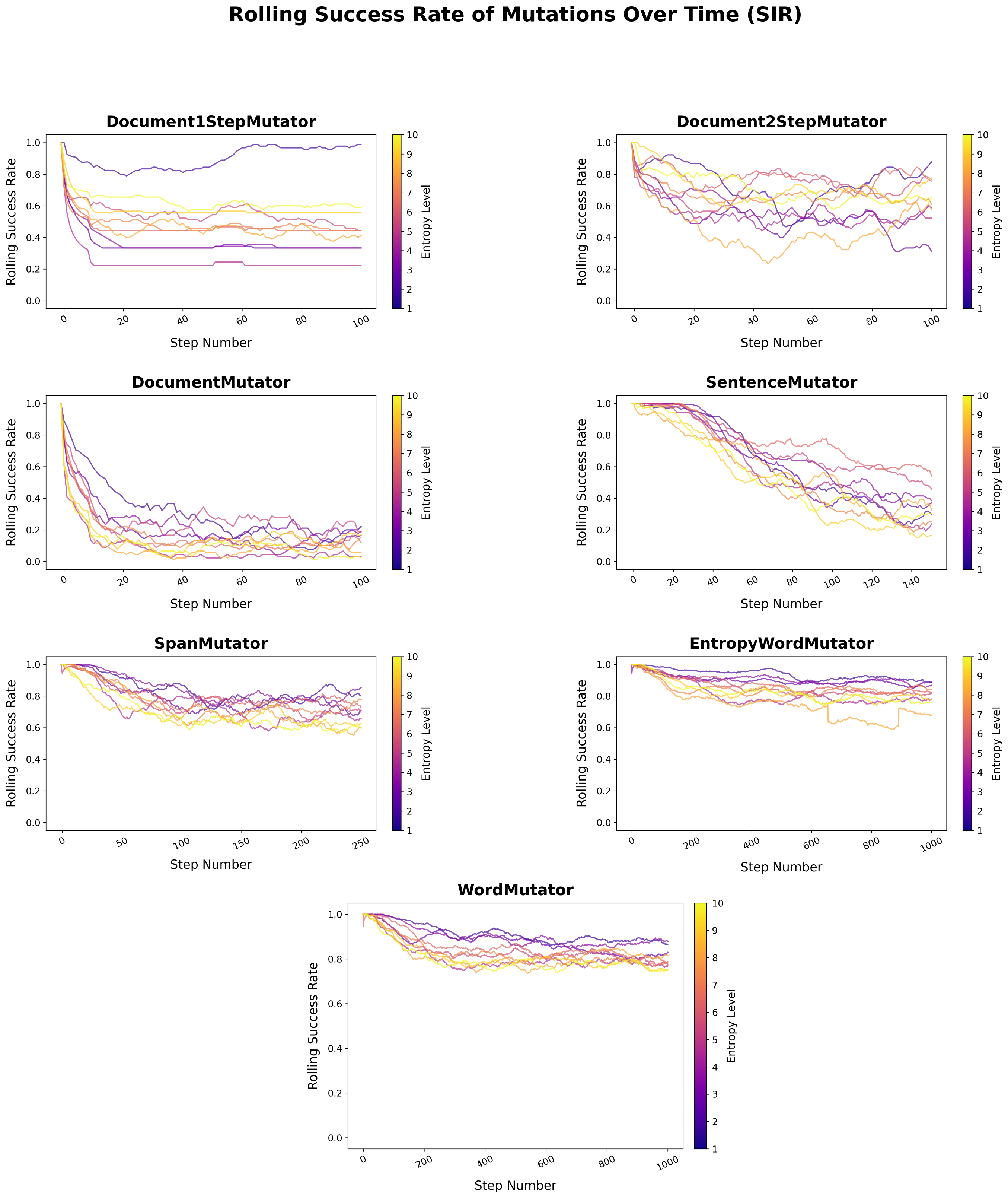}
    \caption{Rolling success rate for the SIR watermark.}
    \label{figure:rolling_success_SIR}
\end{figure}

% Rolling Success Adaptive Figure
\begin{figure}[htbp]
    \centering
    \includegraphics[width=0.95\textwidth]{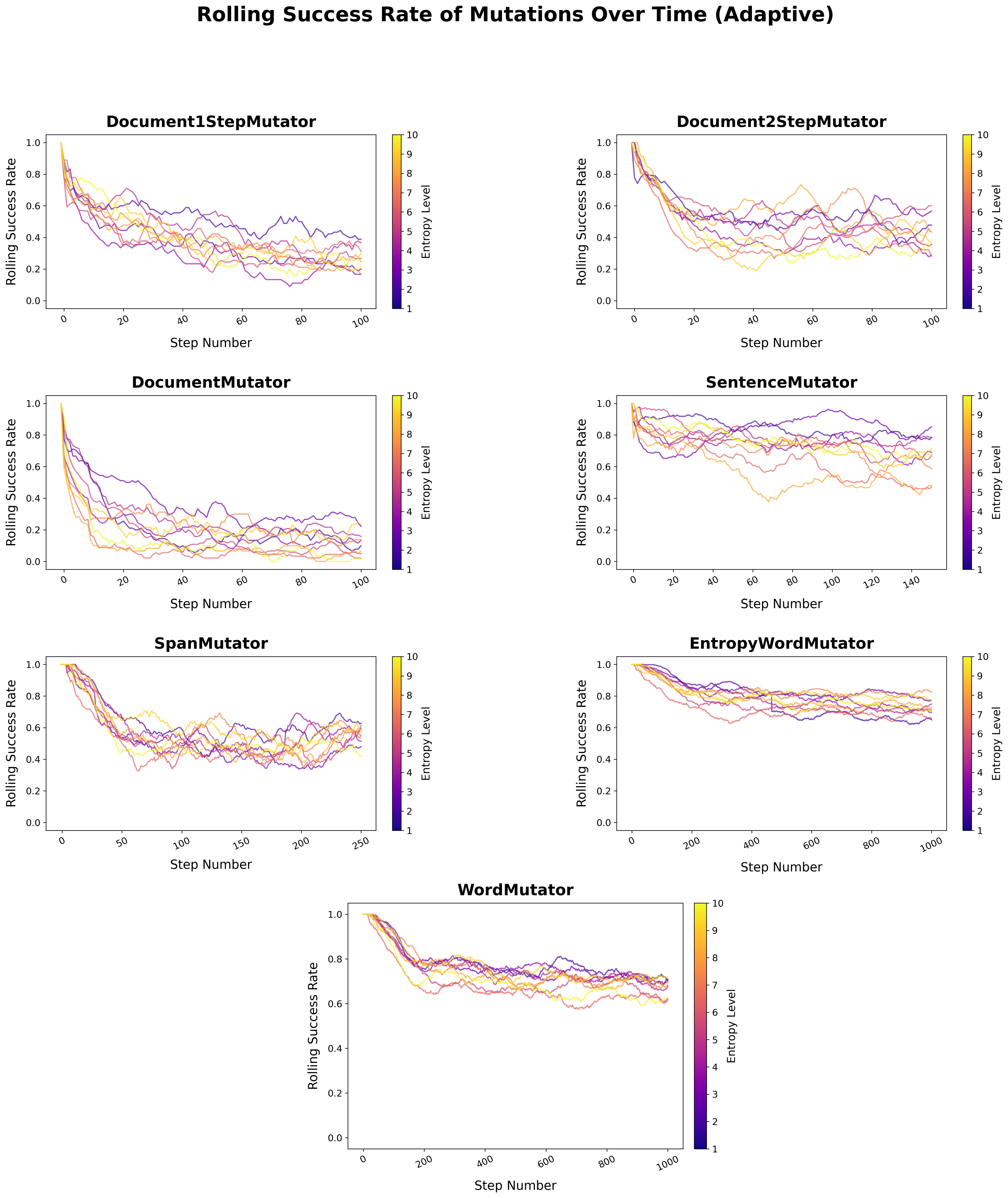}
    \caption{Rolling success rate for the Adaptive watermark.}
    \label{figure:rolling_success_Adaptive}
\end{figure}

\end{document}